\documentclass[electronics,article,submit,moreauthors,pdftex,10pt,a4paper]{mdpi}
\firstpage{1}
\articlenumber{x}
\doinum{10.3390/------}
\pubvolume{xx}
\pubyear{2017}
\copyrightyear{2017}
\externaleditor{Academic Editor: name}
\history{Received: date; Accepted: date; Published: date}

\Title{Emergency Management Systems and Algorithms: a Comprehensive Survey}

\Author{Huibo Bi $^{1}$ and Erol Gelenbe $^{2,}$*}

\AuthorNames{Huibo Bi and Erol Gelenbe}

\address{%
$^{1}$ \quad Beijing Key Laboratory of Traffic Engineering, College of Metropolitan Transportation, Beijing University of Technology, Beijing 100124, China; huibobi@bjut.edu.cn\\
$^{2}$ \quad Intelligent Systems and Networks Group, Department of Electrical and Electronic Engineering, Imperial College London, London SW7 2BT, UK; e.gelenbe@imperial.ac.uk}

\corres{Correspondence: E-Mail: e.gelenbe@imperial.ac.uk; Tel.: +44-207-594-6342; Fax: +44-207-594-6274.}

\abstract{Owing to the increasing frequency and destruction of natural and manmade disasters to modern highly-populated societies,  emergency management, which provides solutions to prevent or address disasters, have drawn considerable research over the last few decades and become a multidisciplinary area. Because of its open and inclusive nature, new technologies always tend to influence, change or even revolutionise this research area. Hence, it is imperative to consolidate the state-of-the-art studies and knowledge to meet the research needs and identify the future research directions. The paper presents a comprehensive and systemic review of the existing research in the field of emergency management from both the system design aspect and algorithm engineering aspect. We begin with the history and evolution of the emergency management research. Then the two main research topics of this area, ``emergency navigation'' and ``emergency search and rescue planning'', are introduced and discussed. Finally, we suggest the emerging challenges and opportunities from system optimisation, evacuee behaviour modelling and optimisation, computing patterns, data analysis, energy and cyber security aspects.}


\begin{document}

\section{Introduction}

Emergency management is the creation of plans and strategies for emergency personals and evacuees to cope with disasters and reduce vulnerability to hazards \cite{drabek1991emergency}. Generally, it can be divided into four aspects, prevention, preparedness, response and recovery (PPRR), which are originated from the work of US State Governors' Association in 1978 \cite{cronstedt2002prevention} and have now been widely accepted as the fundamental components to deal with natural or manmade disasters. The attempts of managing disasters have been deeply rooted in human history and even folklore owing to the impacts of disasters to social and economic aspects of societies. Many great disasters have been recorded in ancient literature such the Bible story of the Deluge. In the ancient times, the operations of emergency management were mostly conducted in a unorganised, reactive manner. Until the 20 century, laws or policies were begun to be passed worldwide to provide financial assistance and investment after or before a disaster strike \cite{haddow2017introduction}, transforming the operations of emergency management to a more organised, proactive fashion. From the 1950s, with the nuclear threat during the Cold War as well as the development of computer technologies, many efforts have been dedicated to computer-aided civil defense programs \cite{Autonomous98,Kaptan}, which motivated the development of the subsequent systematic emergency management studies. The current research efforts in this area can be generally classified into two directions: emergency navigation and emergency search and rescue planning. Emergency navigation, which is also known as emergency evacuation planning, is the process of directing evacuees out of hazardous areas with the aid of real-time routing algorithms or pre-deployed static plans that are based on the prediction and analysis from evacuee behaviour models. Emergency search and rescue planning, on the other hand, focuses on rescuing immobilised and incapacitated evacuees that are trapped in hazardous areas with the assistance of task assignment or resource allocation algorithms. This paper reviews the development and applications of the two main research directions, from both the system design aspect and algorithm engineering aspect. Owing to its open and inclusive nature, new technologies always tend to influence, change or even revolutionise the research area of emergency management. Hence, the current and emerging challenges are also discussed.

The remainder of the paper is organised as follows. We first summarise the research progress in the field of emergency navigation in Section \ref{emergencynavigation}, from the system design aspect in Subsection \ref{emergencynavigation:systems} and algorithm engineering aspect in Subsection \ref{emergencynavigation:algorithms}; then we review the research efforts related to emergency search and rescue planning in Section \ref{emergencysearchandrescueplanning}, including various systems in Subsection \ref{emergencysearchandrescueplanning:systems} and associated algorithms in Subsection \ref{emergencysearchandrescueplanning:algorithms}; next, the research trends and challenges are discussed in Section \ref{challengesandopportunities}; finally, we draw conclusions in Section \ref{conclusions}.

\section{Emergency Navigation}
\label{emergencynavigation}

Nowadays, the research of emergency navigation aims to direct evacuees out of hazardous areas in a timely and safe manner with the assistance of computer-aided systems. In this section, we first review a series of emergency navigation systems chronologically in Subsection \ref{emergencynavigation:systems}, then we summarise various emergency navigation algorithms topic by topic in Subsection \ref{emergencynavigation:algorithms}.

\begin{figure}[!ht]
\centering
\includegraphics[width=0.6\textwidth]{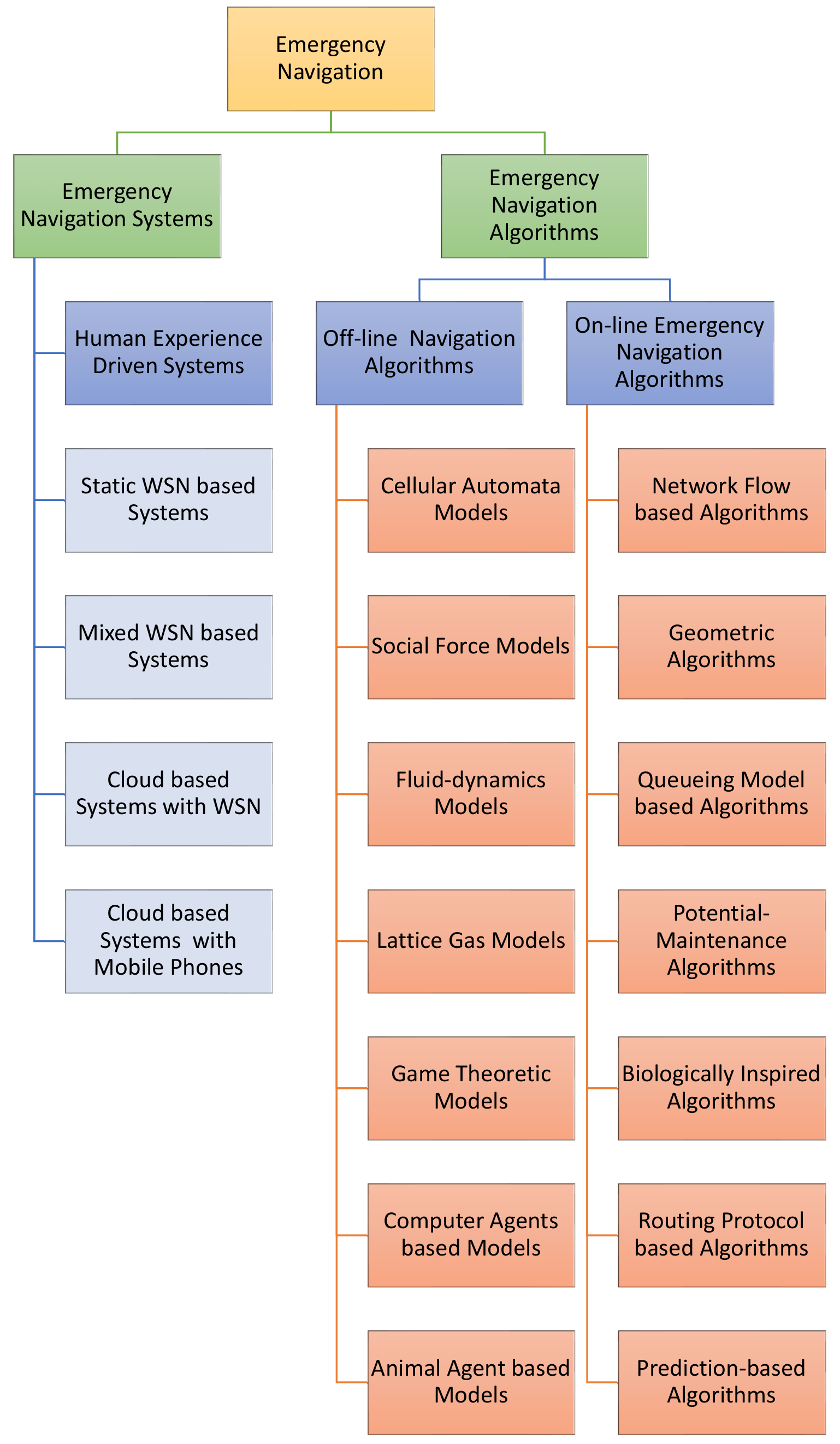}
\caption{A tree diagram explaining the structure of Section \ref{emergencynavigation}.}
\label{fig: EmergencyNavigation}
\end{figure}

\subsection{Emergency Navigation Systems}
\label{emergencynavigation:systems}

The study of emergency evacuation in confined spaces, which was initially motivated by defence applications \cite{Autonomous98,Kaptan}, has attracted much attention owing to the potential of losses in terms of human lives and property during a disaster. Accompanying the advancement of computer technologies, emergency evacuation systems have experienced a few stages: from the original human experience driven systems, to the currently booming in-situ wireless sensor network based navigation systems, towards the cloud based navigation systems which are still in their infancy.

Due to limitations in processing power, early emergency navigation systems are commonly computer-aided information reporting systems to assist emergency managers in making decisions \cite{belardo1984investigation}. Associated emergency navigation algorithms at that time normally used human experience or purely mathematical models to simplify an evacuation process and seek optimal solutions. Before 1990s, research in this area is very limited. The research in \cite{chalmet1982network} considers the evacuation planning problem as a minimum cost network flow problem that converts the original building graph to a time-expanded network; by solving the time-expanded network via a linear programming algorithm, evacuees can obtain optimal routes and achieve shortest evacuation time. The study in \cite{hughes1990graph} designs a graph processing software to represent a underground mine as a graph-based evacuation network in conformity to proper ventilation requirements; each edge is assigned with a weight in terms of its distance to the source of hazard and fresh air intake, and Dijkstra's shortest path algorithm is utilised to find the safest paths for evacuees. The work in \cite{southworth1989road} proposes a traffic monitoring and analysis system to predict the possible traffic jams for emergency planners during urban-scale evacuations; real-time traffic data are collected at roadside traffic counting stations and transmitted to the system via conventional telephone lines; an evacuation simulation model is used to provide the locations and timing of occurrence of potential traffic bottlenecks. The research in \cite{griffith1982hurricane} presents a human experience driven emergency alarm system to facilitate emergency authorities to evacuate residents before the landfall of a hurricane; a ``vertical evacuation'' methodology is proposed to lower the evacuation time and the issuing of ``early evacuation orders'' is believed to be critical in reducing fatalities. To improve the disaster response ability in accidents at nuclear power plants, the study in \cite{zorpette1987planning} proposes an evacuation plan to flee residents within 20 miles of the plant when a radiation leakage occurs. The work in \cite{serpa1981emergency} designs a real-time emergency monitoring and response system for a nuclear power plant; the response decisions are based on discussions between experts at off-site emergency response centers.

With the fast development of information and communications technology (ICT) as well as the advent of low-costing microelectronic devices, in the middle of 1990s, research moved to the development of complex Emergency Cyber-Physical-Human systems to direct evacuees to exits with the aid of an on-site wireless sensor network (WSN). Until today, most of the state-of-the-art emergency navigation systems and algorithms are still based on static WSNs. For instance, the research in \cite{filippoupolitis2009distributed} presents a static WSN based distributed system to compute shortest safe paths for evacuees; this system employs a two-tiered architecture that contains a sensing sub-system and a decision sub-system. The study in \cite{tseng2006wireless} proposes a static WSN based navigation system to direct evacuees with a variant of the temporally ordered routing algorithm \cite{park1997highly}; initially, exit sensors broadcast ``initial packets'' to assign each sensor with an ``altitude'' that is positively correlated with its distance to the nearest exit (sensors that are nearer to the exits possess smaller altitudes than those farther sensors); when a disaster occurs, the altitude of sensors inside a hazardous region will be increased and escape paths will be generated along sensors with higher altitudes to those with lower altitudes. The work in \cite{li2003distributed} utilises a self-organising WSN to guide a robot across a hazardous area; by using an ``artificial potential field'' \cite{koditschek1989robot}, sensors can cooperatively generate a safe path without knowing the network topology.

Compared with static WSNs, mixed WSNs which contain mobile nodes can monitor uncovered areas of static sensors and is less prone to failures in harsh hazardous environments. Hence, some research has employed mixed WSNs to build emergency response systems. For instance, the studies in \cite{gorbil2011opportunistic,gelenbe2012wireless,gorbil2013resilient} have proposed a resilient emergency support system (ESS) with the aid of opportunistic communications \cite{pelusi2006opportunistic}. This system consists of sensor nodes (SNs) and communication nodes (CNs). SNs can detect the hazard in its vicinity and inform the evacuees passing by of the location, while CNs are portable devices that are taken by occupants. The work in \cite{filippoupolitis2011autonomous} presents an indoor autonomous navigation system composed of an intelligent evacuation sub-system (IES) for primary use and a opportunistic communication based evacuation sub-system (OCES) for backup purposes; both sub-systems are supported by pre-installed sensors in the building, the IES utilises static decision nodes to guide evacuees in proximity while the OCES employs mobile decision nodes carried by civilians to disseminate emergency messages and direct evacuees when the IDE malfunctions. The experimental results show that the use of the OCES can considerably reduce the number of fatally injured civilians during an evacuation process. With the increasing ubiquitousness of smart phones which provide powerful sensing ability and suffer less from battery power limitations, many studies have integrated evacuees' portable devices into emergency navigation systems. For instance, the study in \cite{zubair2011adaptive} proposes an emergency support framework that integrates a pre-deployed WSN with the existing mobile network infrastructure to guide evacuees out of a built environment; the framework, namely ``CoWiSMoN'', employs both fixed sensors pre-installed in the building and mobile phones carried by evacuees to collect information and send to a quick rescue response center via short-range wireless communication links or the mobile cellular network; moreover, a cognitive communication protocol that optimises both the network layer and data link layer is designed to ease the network congestion caused by the transmission of large volumes of sensory data and the degradation of communication quality during disasters. Similarly, the research in \cite{inoue2008indoor} presents an indoor emergency evacuation system composed of a sensor-data management sub-system and an indoor navigation sub-system; the sensor-data management sub-system gathers sensory information and can alert users and the building manager via mobile phones; the indoor navigation sub-system utilises radio beacon devices to estimate users' position; each user carries a beacon receiver that can receive signals from beacons and transmit to the mobile phone via Bluetooth.

The major drawback of the WSN based emergency navigation systems is the limited computing capacity, which does not allow them to compute optimal evacuation plans in a timely manner so as to forward this information to evacuees in the presence of time-varying hazards. Hence, some emergency navigation systems have integrated cloud computing technologies that are accessible via on-site WSN, to offload intensive computations to remote cloud servers. For instance, the research in \cite{tseng2007imouse} proposes a hazard surveillance system to detect unusual events in an environment and alert residents; the system is composed of a number of static sensors, several mobile sensors and an external cloud server; when static sensors detect unusually high temperatures, mobile sensors will be dispatched by the cloud server to take snapshots and upload to the server for further analysis; if a fire emergency is confirmed, the cloud server will notify the residents in the vicinity to evacuate. the study in \cite{dong2014multicloud} proposes a multi-cloud based evacuation system that integrates an on-site WSN and remote cloud servers to calculate evacuation paths for users; when a disaster occurs, the system launches an instance for each user to compute the desired evacuation route; owing to the limited I/O capability of cloud providers, several cloud platforms are employed to launch sufficient instances for evacuees and a dynamic programming algorithm is used to minimise the overall latency and service maintenance cost of the system.

Although the hybrid emergency response systems that integrate on-site WSNs and off-site cloud servers can avoid the problems caused by the limited computing power of WSNs, the disadvantages of limited battery life-time of the WSNs, as well as the high likelihood of system malfunction during an emergency still remain. Hence, many of the studies have replaced WSNs with smart phones carried by evacuees to build more flexible systems. For example, the work in \cite{chu2011integrated} presents a building fire evacuation system that consists of radio frequency identification (RFID) sensor tags, mobile phones with RFID reader and a back-end cloud server; since signals from the global positioning system are unavailable inside built environments, RFID sensor tags are used to record the temperature and location information; when a fire breaks out, mobile phones carried by evacuees will periodically sense the RFID signals and upload to the cloud server; the cloud server will then calculate the shortest safe route for each civilian with respect to the distance to exit and the summed temperature along the route. The research in \cite{qiu2014cloud} proposes a smart cloud evacuation system (SCES) to post emergency messages and plans to residents in a built environment; in the front end, a wireless intelligent sensor network (WISN) that integrates a WSN with smartphones are utilised to collect information; in the back end, a cloud based decision-making system is used to analyse the uploaded multimedia information (such as voice, text, images, etc.) and calculate escape paths with a 3D simulator. The study in \cite{gelenbebi2014Emergency} proposes an infrastructure-less emergency navigation system to guide evacuees out of hazardous buildings with the aid of smart phones carried by evacuees and an off-site cloud based decision support system (CDSS); evacuees can locate themselves by taking a snapshot of pre-deployed landmarks (e.g. door signs) in the vicinity and uploading it to the CDSS for location identification; the CDSS computes congestion-aware paths with the shortest time to exits for evacuees based on their uploaded locations; evacuees are guided to exits in loose groups with the assistance of a combination of the social potential fields algorithm and a cognitive packet network based algorithm; to reduce the likelihood for the battery power of smart phones to be drained during the energy-hungry communication among smart phones and the CDSS, a power-aware communication protocol is also presented to balance the remaining battery power of smart phones by relaying sensory information via more energy efficient short-range communication techniques before uploading to the cloud server through 3G.

\begin{table}[h!]
    \begin{center}
        \begin{tabular}{| r | c |}
        System type  &   Period \\
        Human experience driven systems  & 1970s - 1990s \\
        Static WSN based systems   & 1995 - present \\
        Mixed WSN based systems   & 2006 - present \\
        WSN \& Cloud based systems  & 2007 - present \\
        Cloud based systems  \& mobile phones & 2011 - present \\
        \end{tabular}
        \caption{Emergency Navigation Systems}
        \label{table: sysmbols}
    \end{center}
\end{table}

\subsection{Emergency Navigation Algorithms}
\label{emergencynavigation:algorithms}

As the kernel of an emergency navigation system, many studies have concentrated on emergency navigation algorithms, which aim to guide evacuees out of hazardous areas safely and efficiently. Previous emergency navigation algorithms can be divided into two types: off-line algorithms and on-line algorithms. Off-line algorithms focus on optimising the design of crowded sites and evaluating the overall clearance time for all evacuees before a disaster occurs. On the other hand, on-line algorithms aim to provide evacuation paths for evacuees in a real time manner. The literature review for the two categories of algorithms is detailed as follows.

\subsubsection{Off-line Emergency Navigation Algorithms}

Since research has indicated that destructive crowd behaviours, such as clogging, pushing and trampling, can lead to serious fatalities \cite{Helbing1}, also owing to the absence of real evacuation data \cite{yang2005simulation}, off-line emergency navigation algorithms have been dedicated to investigate and design crowd behaviour models \cite{gwynne1999review,zheng2009modeling} to simulate the crowd movements in reality and prevent destructive crowd behaviours from occurring by improving the design of built environments. The crowd behaviour models of the off-line emergency navigation algorithms can be classified into cellular automata models \cite{wolfram1983statistical,Bandini1,yu2007cellular,spartalis2014crowd,muller2014study}, social force models \cite{helbing1995social,Helbing1,parisi2005microscopic,seyfried2006basics}, fluid-dynamics models \cite{henderson1971statistics,helbing2002simulation}, lattice gas models \cite{fredkin2002conservative,tajima2001scaling,tajima2001scaling,takimoto2003spatio}, game theoretic models \cite{hoogendoorn2003simulation,lo2006game,ehtamo2010modeling,zheng2011modeling}, computer agents based models \cite{bonabeau2002agent,zarboutis2004searching,goldstone2005computational,pan2007multi} and animal agent based models \cite{saloma2003self,altshuler2005symmetry}.

Cellular automata models discretise a given structure into uniform ``cells'' that each cell can hold one person. This approach can precisely model the influence of an individual's physical dimensions, but is ineffective in depicting the movement speed and direction, due to the discrete spatial structure \cite{pelechano2008evacuation}. The physical conditions and the movement patterns of evacuees are normally determined by a set of local rules at each cell (one drawback is that it is relatively difficult to customise the physical attributes of each individual civilian). Since this model can effectively mimic the interactions between the environments and the pedestrians, many studies have utilised this microscopic model to simulate the pedestrian dynamics during evacuations in the last two decades. In these models, evacuees are considered as either homogeneous with identical physical attributes (e.g. gender, age, mobility, psychology) or heterogeneous individuals with different characteristics. For instance, the research in \cite{perez2002streaming} utilises a homogeneous cellular automata model to investigate the exit dynamics of evacuees in a room with different number of exits; the arching behaviour, which is a signature of jamming that happens when the exits are overused, is observed near the exits; a ``power law behaviour'' is also found: when the exit door can evacuate more than one evacuee at a time, the evacuees will escape from the room in bursts of various sizes. In \cite{spartalis2014crowd} a heterogeneous cellular automata model mimics the evacuation process in a retirement house; evacuees initially belong to three groups (middle-aged people, nursing staff and older people), and groups are also formed dynamically due to the follow-the-leader effect. In \cite{muller2014study} grouping behaviours in evacuations are induced by introducing ``bosons'' into cells of the floor field cellular automaton \cite{Burstedde1}; bosons are placed by evacuees as markers to increase the probability for other group members to reach some particular cells. The resulting simulations indicate that the evacuation time decreases with the increasing numbers of groups.

The research in \cite{helbing1995social} first proposed that the motion of a crowd of pedestrians are subject to ``social forces''; in the social force model, the motion of a pedestrian is mainly affected by the destination, the repulsive forces from other objects (e.g. the pedestrian keeps a certain distance away from other pedestrians or obstacles), the attractive forces from other objects (e.g.  the pedestrian is attracted by friends or window displays); a ``direction dependent weight'' is introduced into the model since the objects behind a pedestrian have a weaker effect on the pedestrian; a ``fluctuation effect'' is also integrated into the model to simulate the random movement behaviours or deliberate deviations from usual motion rules. The study in \cite{zheng2002collective} combines the social force model with a counterpropagation neural network model \cite{hecht1988applications} to mimic crowed behaviours in panic; the personality of evacuees is expressed as impatient and patient. the velocity and the action of evacuees is determined by the social force model and the neural network model, respectively; the neural network has four inputs: the personality of an evacuee, the deviation between the desired speed and real speed, the space on the left side of the evacuee and the space on the right side of the evacuee; the output of the network is the action of the evacuee: follow the person in front, evade the front person from the left side and evade the front person from the right side. The research in \cite{parisi2005microscopic} utilises the social force model to investigate the pedestrian evacuation dynamics in a room with an exit; experimental results show that if evacuees move at the low desired velocities, the faster the evacuees moves, the faster they will evacuate the room; however, if evacuees move at the high desired velocities, the ``faster is slower'' effect, that the faster the evacuees wish to move the slower they can escape from the room, is observed and analysed.

Fluid-dynamics models imitate evacuee flows as fluids to study the density and speed adaptation during an evacuation process. Compared with microscopic models, the macroscopic fluid-dynamics models are better at simulating and analysing the behaviours of large crowds. For instance, the research in \cite{hughes2002continuum} derives several equations that govern the motion of a pedestrian flow from the ``continuity equation'' of fluid mechanics in physics; the proposed equations lead to two possible regimes of a pedestrian flow: the fast-moving, low-density ``supercritical'' flow in which disturbances spread within the flow and the slow-moving high density ``subcritical'' flow in which disturbances are swept along by the flow; several partial differential equations, that govern the crowd behaviours of a flow that contains different types of pedestrian, are also studied; a pedestrian type is determined by the destination, walking speed and perception; the analysis and experimental results show that pedestrians tend to reach each immediate destination in minimum time rather than arriving at all destinations in overall minimum time. The study in \cite{colombo2005pedestrian} presents a continuum model to investigate the relation between evacuee density and walking speed during a process of evacuees leaving a corridor through a door; the proposed model is based on the Lighthill-Whitam \cite{lighthill1955kinematic} and Richards \cite{liu1975riemann} model that is used to simulate vehicle flows; specifically, this model describes the decrease in the outflow through a door caused by the panic ``overcompression'' effect of evacuees in front of the door; the analytical results indicate that the rise of panic can dramatically decrease the outflow of evacuees when the door is narrow.

Since a pedestrian flow is a many-body system \cite{thouless2013quantum} that is composed of strongly interacting persons, lattice gas models that consider pedestrians as particles on the square lattices have attracted attention since 1980s. The research in \cite{tajima2001scaling} utilises a lattice gas model to simulate the process of a pedestrian flow evacuating a hall; the hall is represented by square lattices and evacuees are randomly distributed over the lattices; each evacuee can either hold still or move in four directions: forward, backward, left and right; evacuees move in the preferential direction with no back step and cannot overlap on lattices occupied by other evacuees; different dynamical patterns such as arching, flattening and pitting are observed in computer simulations; experimental results indicate that the dynamical phase transition from the choking flow to the decaying flow occurs at a critical time. The study in \cite{takimoto2003spatio} employs a lattice gas model to study the evacuation time for a crowd to escape from a hall through a single exit; evacuees are modelled as biased-random walkers and move in preferential directions; the hall is represented by square lattices and each square lattice may contain up to one evacuee at a time; the spatio-temporal distribution of evacuation time of evacuees is derived from simulations; the experimental results show that the evacuation time of an evacuee depends highly on its initial position within the hall; the effect of the exit width, initial population density and urgency level are also investigated in the experiments.

To explicitly model the behavioral reactions of the individuals during an evacuation process, especially the cooperative and competitive behaviours \cite{kirchner2003simulation}, many of the studies have utilised game theory to mimic the interactive decision-making and strategy-adapting among evacuees. For instance, the research in \cite{lo2006game} employs the non-cooperative game theory \cite{fudenberg1991game} to mimic evacuees' exit selection process when an emergency occurs in a building with multiply exits; the procedure of the algorithm consists of two steps; in the first step, all evacuees are considered as a ``whole entity'' which aims at minimising overall evacuation time while a ``virtual entity'' is used to maximise the overall evacuation time by imposing the blockage influence; hence, a two-player zero sum game is envisaged between the evacuees and the virtual entity; the optimal strategy is found when a Nash Equilibrium \cite{nash1951non} is achieved via optimising the probabilities for the evacuees to choose each exit and the possibilities for the virtual entity to pick each exit (to generate congestion); in the second step, the decision of each individual evacuee is determined by calibrating the evacuees' probabilistic choices in terms of evacuees' distance to exits; this is because, in reality, an evacuee will not pick a farther exit unless the nearer exit is congested. The study in \cite{zheng2011modeling} utilises a game-theoretical model to investigate the competitive and cooperative behaviours during an evacuation process from a large single room with one exit; During the evacuation process, when $N$ evacuees wish to occupy the same desired position, the conflict among evacuees leads to a $N * N$ game; each evacuee can choose to either compete or cooperate: if all the evacuees choose to cooperate, then they will all reach the desired position; if all the evacuees are competitive, they will all be blocked at the initial position; if one evacuee choose to compete and the rest is in a state of cooperation, only the competitive evacuee can reach the desired position; the simulation results show that: $(1)$ with the increasing urgency of emergency, the cooperation among evacuees decreases; $(2)$ higher cooperation frequency will result in shorter overall evacuation time; $(3)$ the imitation behaviours among evacuees can enhance the cooperation level but reduce the efficiency of the evacuation process.

Algorithms based on pure mathematical models have difficulty in fully representing and capturing the dynamics of an evacuation process. Hence, the agent-based algorithms, which normally represent a hazardous environment with a number of autonomous decision-making virtual agents, have drawn considerable attentions in recent years. One major advantage of the agent-based algorithms is the ability to evolve and learn, which can lead to unanticipated behaviours during simulations. This characteristic makes the agent-based algorithms a canonical approach to mimic the counterintuitive emergent phenomena \cite{bonabeau2002agent}. For instance, the research in \cite{zarboutis2004searching} utilises a multi-agent framework to simulate a metro system in the case of a tunnel fire; the passengers and metro personnel, the technological system, as well as the fire and smoke are simulated by separate agents and co-evolve in an interactive manner; an effective evacuation plan is designed by varying environmental factors, such as the number of passengers on the train, the time cost for the train driver to open the doors, etc.; with the aid of the multi-agent computer simulations which can test different scenarios, the emergency personnel can quickly customise a rescue plan when a disaster occurs. The study in \cite{pan2007multi} presents a prototype multi-agent simulation system that can build a virtual environment with autonomous agents for safe egress analysis; the proposed system consists of a geometric engine that represent the physical environment with AutoCAD, a population generator that can produce evacuee agents with diverse age, mobility, etc., a global database which maintains all the state information of agents; an events recorder that captures the behaviours of evacuee agents, a visualiser which displays the movement of evacuees, a crowd simulation engine that is assigned to each evacuee agent to manage the individual behaviour in terms of the perception-action approach \cite{nakamura1995motion}; each evacuee agent is modelled to makes decisions based on three basic conventions: instinct, experience and bounded rationality \cite{march1994primer}; some emergent behaviours such as competitive, queueing and herding are observed in the simulation.

Owning to the scarcity of human emergent behavioural data and the difficulty in conducting genuine emergency evacuation experiments, the studies of emergent behaviours have largely depended on simulations. Hence, in recent years, animals have been used in escape panic experiments to study crowd evacuation. For example, the research in \cite{saloma2003self} employs mice to study an indoor evacuation process; mice were released into a rectangular container (simulate a large single room) filled with tap water and were left to swim towards a dry platform; an exit is placed between the wet and dry areas to simulate the door of the single room; the effect of exit width and exit number on the mouse escape rate is investigated over different experimental sessions, which are recorded with a digital video camera; the experiments demonstrate some well-known behaviours of panicking crowd: when the exit width approximately equals to the size of a mouse, the diffusive evacuation flow is observed; when the exit width becomes larger, the mice evacuate the exit in bursts of different sizes that yield the power-law distributions depending on the exit width. The work in \cite{altshuler2005symmetry} employs a species of Cuban leafcutter ants called \emph{Atta insularis} to investigate the effect of panic-induced herding to an evacuation process from a two-exit room; ants are introduced into a circular acrylic cell with two exits symmetrically situated at left and right; in the first experiment which simulates a normal evacuation process, when the ants are placed into the cell, the two exits are opened synchronously so that the ants can escape; in the second experiment which mimics an emergency evacuation process, the only difference from the first experiment is to inject a dose of insect-repelling liquid to generate a panic before opening the exits; the experimental results show that ants escape from both exits in approximately equal proportions in normal conditions but prefer one of the exits in emergency conditions; the experiments demonstrate the theoretical prediction that the herding behaviour in confined spaces can generate a non-symmetrical use of two identical exit doors; in addition, the observed evacuation dynamics are reproduced with a computer model inspired by \cite{Helbing1}.

\subsubsection{On-line Emergency Navigation Algorithms}
\label{emergencynavigationalgorithms: onlinealgorithms}

Contrary to off-line emergency navigation algorithms which aim at optimising the design of crowded sites or generating evacuation plans for facility managers via developing various crowd behaviour models and computer simulations, on-line emergency navigation concentrates on combining mathematical models \cite{kuligowski2005review} or algorithms \cite{GelenbeW12,GelenbeW13} with underlying sensing, communication and computational devices to guide evacuees out of hazardous environments in a real time manner.

Since on-line emergency navigation algorithms require real time information exchanges with the hazardous environment, these algorithms are usually integrated into emergency navigation systems. With the development of emergency navigation systems which are detailed in \ref{emergencynavigation:systems}, various emergency navigation algorithms have been proposed such as network flow based algorithms \cite{chalmet1982network,francis1984negative,kisko1985evacnet,lu2003evacuation,lu2005capacity}, geometric algorithms \cite{chen2008distributed,wang2013sensor}, queueing model based algorithms \cite{macgregor1991state,cruz2005service,stepanov2009multi,lino2009modeling,desmet2013graph,huibo2016flowoptimisation,bi2016energyaware}, potential-maintenance algorithms \cite{li2003distributed,tseng2006wireless,chen2008load}, biologically inspired algorithms \cite{RNN2008,jankowska2009wireless,li2010multiobjective}, routing protocol based algorithms \cite{filippoupolitis2010emergency,gelenbe2012wireless,BiDesmetGelenbeISCIS2013,biandgelenberouting2014} and prediction-based algorithms \cite{hasofer2001stochastic,barnes2007emergency,han2010firegrid,radianti2015spatio,BIGELENBEIEEEPERCOM2015}.

Network flow based algorithms consider the evacuation planning problem as a minimum cost network flow problem \cite{ford2010flows,ahuja1988network}. Commonly, this type of algorithm first predicts the upper bound of the overall evacuation time and then convert the original building model to a time-expanded network by duplicating the original network for each discrete time unit. After that, linear programming or heuristic algorithms are utilised to compute the optimal evacuation plan. This type of approach can achieve the optimal solution but normally does not take the spreading of the hazard into consideration. For instance, the work in \cite{chalmet1982network} utilises a dynamic network optimisation model to minimise the overall evacuation time and prevent ``bottlenecks'' from occurring in a large building; the building is represented by a graph model composed of nodes and arcs; the capacity of a node is determined by dividing the space area of the node by the typical space occupied by an evacuee; the capacity of an arc, which is defined as the maximum number of evacuees that are allowed to traverse the arc per unit time, is determined by the passageway width; the graph model is expanded into a time-expanded network by duplicating the original graph model over $T$ time periods, where $T$ is determined by dividing the approximate evacuation time $T_e$ by the length of a time period (10 seconds); to reduce the computational complexity and ensure the existence of a feasible solution, the minimum feasible building evacuation time $T_e$ is determined by the proposed bisection search algorithm; the time-expanded network is solved via a large-scale dynamic transshipment algorithm from the GNET program  \cite{graves1977design}. Since the search complexity of a time-expanded graph grows exponentially with the increase of the time bound $T$, the studies in \cite{hoppe1994polynomial,hoppe2000quickest} develop a polynomial time algorithm to solve the evacuation problem with a fixed number of sources and exits; the evacuation problem is converted to a quickest flow problem, which aims to send a specific amount of flows from sources to sinks in the shortest time; the building model is represented by a graph with integral transit times and capacities on the edges; the evacuees flows are represented by the temporally repeated flows proposed in \cite{ford2010flows} rather than static flows in a time-expanded network; the quickest flow problem with multiple sources and sinks is then transferred to a lexicographic maximum dynamic flow problem and can be solved by using the algorithms presented in \cite{megiddo1974optimal} and \cite{minieka1973maximal}. Since linear programming algorithms that utilise time-expanded networks to calculate optimal evacuation plans can suffer from high computational cost, the work in \cite{lu2003evacuation,lu2005capacity} propose a heuristic-based algorithm called capacity constrained route planner (CCRP) to produce sub-optimal evacuation plans in a time-efficient manner; rather than transforming the original evacuation network into a time-expanded network, CCRP employs the Dijkstra's shortest path algorithm \cite{dijkstra1959note} to search only the original evacuation network and calculate the quickest routes for evacuees; CCRP first searches the route with the shortest arrival time from any source node to any destination node in terms of path length, previous reservations and possible waiting time; then it allocates evacuees to this route with respect to the capacity of the route; the CCRP algorithm will iterate the above two steps until all the evacuees reach the exits. These approaches can theoretically solve optimal routes with the shortest time to exits by avoiding congestion. However, to achieve shortest time to exit, evacuees must accurately follow the suggested paths and reach every node on schedule and may even wait certain time at a node to avoid congestion. This is impractical in a real evacuation process. Moreover, these approaches suffer from high computational complexity because the time-expanded network will contain at least $(N+1)T$ nodes for a graph with $N$ nodes and an upper bound of evacuation time $T$. In addition, as aforementioned, the spreading of the hazard is not considered in these approaches.

Geometric algorithms normally use a graph model to represent a hazardous environment and take advantage of the unique properties of geometric graphs to calculate safe egress paths for evacuees. For instance, the research in \cite{chen2008distributed} adopts the localized Delaunay Triangulation method \cite{li2002distributed,li2003localized} to partition a wireless sensor network into triangular areas and construct area-to-area egress paths with the aid of a distributed navigation protocol; each sensor, which is the shared vertex of all the adjacent triangles, maintains the node ID, hops to the exits and the sensed hazard level (temperature) of the neighbour sensors; the direction of an egress path is generated from vertices with larger hop count to the exit to vertices with smaller hop count; the safety level of a triangle area is classified into three color-coded levels (red ``high'', yellow ``moderate'' and green ``low'') by comparing the average detected temperature of the associated sensors with a pre-set temperature threshold; in built environments with multiple-exits, additional wireless access points (AP) that can cover the whole environment are deployed in the vicinity of each exit to count the number of evacuees nearby, a load-dispersion algorithm is employed to distribute evacuees by limiting the number of users per exit. The research in \cite{wang2013sensor} proposes a WSN based emergency navigation system to guide evacuees without the aid of any pre-knowledge of sensor or user locations; the process of navigating evacuees to the exit contains three stage: firstly, a road map is generated as the backbone route; secondly, the exit is connected to the backbone route via a virtual power field algorithm; thirdly, evacuees are directed to the backbone route via the virtual power field algorithm and then follow the backbone route all the way to the exit; the road map is constructed via concatenating the medial axis of the boundary of any two safe areas; as is proven in \cite{bruck2007map}, the medial axes of the safe regions, which can form continuous curve graphs, retain the topological and geometric features of the safe areas; in the virtual power field algorithm, the virtual power of a point is inversely proportional to its distance from the hazard; the route from any point to the backbone route will follow the most descending  direction of the virtual power field; owing to the expanding or shrinking of the hazard, the dangerous areas vary during the evacuation process; hence, a local road map updating algorithm is proposed to rebuild the backbone route of the affected areas instead of reconstructing the entire backbone route when a variation of the dangerous areas is detected. However, the effectiveness of these approaches highly depends on the topology of the deployed wireless sensor network. The change of the topology will induce redeployment and re-calibration of these algorithms.

Owing to the stochastic, highly transient and nonlinear nature of an evacuation process, queueing models have been proven as a useful tool to capture and analyse the dynamics of evacuees. Normally, by treating significant locations such as doorways or staircases as ``servers'', queueing model based approaches \cite{Muntz}, which generalise the Markovian models of computer systems \cite{Unified}, transfer building graphs to a queueing network or a number of isolated ``queues'' to estimate congestion and evacuation delays. For instance, the process of pedestrians traversing a corridor or stairwell is analysed as a state-dependent process in \cite{macgregor1991state}, a $M/G/C/C$ state-dependent queue model is utilised to estimate the congestion delays at corridors or stairwells and the overall evacuation time of an evacuation process; the pedestrian flows are classified into three categories: uni-directional flow, bi-directional flow and multi-directional flow; the relationship between the crowd density and the mean walking velocity of evacuees in the three categories of pedestrian flows are derived from \cite{fruin1971pedestrian}; the capacity of a corridor or stairwell is calculated based on \cite{tregenza1976design}, which indicates that the evacuee flow will cease to move when the population density reaches 5 evacuees per square meter; the state-dependent service rate of the three categories of pedestrian flows can be calculated in terms of the mean walking speed and the corridor capacity; finally, the time cost for an evacuee flow to traverse a corridor or a stairwell can be computed by the mean value analysis (MVA) algorithm introduced in \cite{francis1980network}. To ensure no corridors will be block during an evacuation process in a built environment, the work in \cite{cruz2005service} considers the evacuation planning problem as a service and capacity allocation (SCA) problem and searches the smallest capacity of each corridor via modelling the building as a $M/G/c/c$ queueing network; the $M/G/c/c$ queueing network is employed to calculate the average queue length at each corridor with the following steps: $(1)$ the average walking speed $V_n$ of $n$ evacuees in a corridor is calculated by the equations derived from the congestion model proposed in \cite{yuhaski1989modeling}, $(2)$ the state-dependent service rate $f(n)$ with $n$ evacuees in a corridor can be computed by $f(n) = \frac{V_n}{V_1}$, where $V_1$ is the average speed of a lone evacuee, $(3)$ term $p_n$, which is the probability of $n$ evacuees in a corridor can be calculated by the equations derived from \cite{cheah1994generalized}, $(4)$ the average queue length of a corridor can then be computed by $L = \sum_{n = 1}^{c} np_n$; to analyse the smallest capacity of each individual corridor, the generalised expansion method \cite{kerbachea1987generalized,kerbache1988asymptotic} is used to expand the $M/G/c/c$ queueing network into an equivalent Jackson network via adding an artificial holding node in front of each finite queue to register the blocked evacuees due to capacity limitation; After decomposing the queueing network, a local search algorithm inspired by  \cite{smith2000performance} is used to search the smallest feasible capacity of each queue. Similarly, the studies in \cite{lino2009modeling,lino2011tuning} utilise a $M/G/c/c$ queue model to simulate the dynamics and predict the overall evacuation time of an egress process without hazard; rooms, corridors and stairways are modelled as queues in which the service rate depends on the evacuee density; doors, exits and gateways are imitated as queues in which the service rate depends on not only the evacuee density, but also the faster-is-slower effect \cite{helbing2000simulating} and the crowd impatience \cite{wang2008modeling}; to validate the effectiveness of the queue system, a discrete-event simulation model is implemented via the SimEvents toolbox in the MATLAB/Simulink environment and experimental results show that the egress time of evacuees in simulations highly matches with the prediction of the proposed queueing model. Rather than simulating all the building components as $M/G/c/c$ queues, the research in \cite{watts1987computer} models doorways that can pass one person at a time as $M/M/1$ queues; on the other hand, corridors or stairs are modelled as $M/G/\infty$ queues, in which the infinite number of servers imply that no congestion occurs in corridors or stairs. Rather than using traditional closed network models which suffer from high computational costs, the study in \cite{desmet2013graph} proposes a computationally efficient open network model with product form to predict the congestion level at each point of interest (PoI) and the overall evacuation time with respect to average arrival and departure rates at each observation point; By assuming Poisson arrivals of evacuees at each originating location, uni-directional corridors that allow at most one evacuee to pass at a time and exponentially traversal delays at each corridor, a $M/M/1$ queue model is established to mimic each corridor; hence, the average delay at a corridor can be calculated by $\frac{1}{\mu - \lambda}$, where $\frac{1}{\mu}$ represents the average traversal time of a corridor and $\lambda$ represents the average arrival rate of evacuees at a corridor; the average traversal time of a path can be calculated by summing the average delay of each corridor on it. Rather than considering each significant location (such as a doorway or staircase) as an independent ``queue'' and then use either the limiting probabilities for the number of customers in an M/G/C/C state-dependent queueing model \cite{cruz2005m} or steady-state solutions \cite{desmet2013graph} to analyse the number of evacuees at the location, the work in \cite{huibo2016flowoptimisation} treats all the significant locations in the designated area as a ``queueing network'' by considering the interaction effects of various evacuees among linked ``queues''; in this study, to predict the time cost $T$ for an evacuee to traverse a path, a G-network model \cite{gelenbe1998multiple} is employed to periodically compute the utilisation rate of each node and edge under the combined impact of a specific routing scheme and panic behaviours; Little's formula is then used to calculate the average delay of each node and edge; finally, $T$ can be calculated by summing the average latency of each node and edge on it. The research in \cite{bi2016energyaware} proposes an urban scale emergency navigation system to guide vehicles to safe zones in the aftermath of a disaster in a latency and energy efficient manner; a G-network model \cite{gelenbe1993g} is utilised to analyse and capture the dynamics of vehicles under the joint influence of interactions among individual vehicles and the re-routing decisions from the navigation system; by using this G-network model, the average number of vehicles and the average traversal time at each intersection or road segment can be calculated; hence, the total average delay experienced by a vehicle and the total fuel consumption in the network can be described by a goal function; finally, a gradient descent algorithm is utilised to reduce the time and fuel cost (minimise the goal function) by optimising the probabilistic choices of linked road segments at each intersection.

Potential based algorithms normally can dynamically develop navigation paths by assigning attractive or repulsive potentials to the exits and hazards, and the evacuees move as a result of the net attraction-repulsion in various directions. For instance, the research in \cite{li2003distributed} presents a self-organizing sensor network to guide users such as robots, evacuees or unmanned vehicles out of a hazardous environment along safest paths by using the ``artificial potential fields'' algorithm \cite{koditschek1989robot}: when a sensor detects hazard, it will broadcast emergent messages including sensor ID, number of hops to the arrived sensor ($N_h$) to other sensors; when a sensor receives multiple emergent messages from the same hazardous sensor, it will keep the smallest $N_h$; the potential value of a sensor generated by a hazardous sensor is calculated by $\frac{1}{N_h^2}$; hence, the overall potential value of the sensor is computed by summing the potential value generated by each hazardous sensor; in this way, an attractive force is generated by the destination to pull the user while repulsive forces are generated by the dangerous zones to push the user away from them; the safest path is generated by following the most descending direction of the potential field; experiments are conducted on a testbed with 50 Mote MOT300 sensors \cite{hill2000system} and the results indicate that the algorithm could successfully direct the objects to the destination; however, multiple destinations may have a negative impact on the efficiency of reaching the exits as the users move under the actuation of artificial forces; Moreover, the convergence time for network stabilization is relatively long due to the effect of data loss, asymmetric connection and network congestion. The study in \cite{tseng2006wireless} proposes a temporally ordered routing algorithm \cite{park1997highly} based multi-path routing protocol to route evacuees to exits through safest paths; a navigation map is manually defined during the deployment process to avoid impractical paths; in the initialisation phase, each sensor is assigned with an altitude with respect to its hops to the nearest exit: sensors nearer to the exits are assigned with smaller altitudes while sensor farther from the exits are allocated with larger altitudes; when an emergency event is detected, a sensor $s_i$ within the hazardous regions will update their altitudes by $A'(s_i) = max \big\{ A(s_i), A_{emg} \times \frac{1}{h^2_{s_i,s_h}} + h_{s_i,s_e} \big\}$, where $A'(s_i)$ and $A(s_i)$ represent the altitude of sensor $s_i$ before and after update; term $A_{emg}$ is a large constant, term $h_{s_i,s_h}$ represents the shortest hop distance between sensor $s_i$ and the hazardous sensor $s_h$ while term $h_{s_i,s_e}$ represents the shortest hop distance between sensor $s_i$ and the exit $s_e$; a hazardous region is constructed by sensors within a predefined hop distance from the hazardous sensor; egress routes are generated from sensors with higher altitudes to senors with lower altitudes; therefore, the update of altitudes of hazardous sensors can ensure evacuees bypass the hazardous regions. The work in \cite{pan2006emergency} extends the algorithm in \cite{tseng2006wireless} to a 3D environment and divides the sensors into normal sensors, exit sensors and stair sensors in terms of location; if no available path to exits can be discovered, evacuees will be directed to rooftops and wait for rescue. However, multiple destinations (exits) may affect the efficiency of reaching the exits as the users move under the actuation of artificial forces. Moreover, the convergence time for network stabilization is relatively long due to the effect of the information synchronization, asymmetric connection and network congestion.

Millions of years of evolution has made the animal foraging behaviours become near-optimal solutions of autonomous search and path-finding \cite{gelenbe1997autonomous}, biologically inspired approaches, which are inspired by simple but reliable natural mechanisms, employ heuristics to search optimal routes in a computationally efficient manner. For instance, a feed-forward neural network model is adopted to a wireless sensor-actuator network (WSAN) for evacuation routing in \cite{jankowska2009wireless}; all physical nodes in the WSAN deploy a neural network with identical topology: an input layer, a hidden layer and an output layer; the input layer receives the latest two coordinates of a pedestrian and a suggested direction is subsequently generated by the output layer; the neural networks are trained with a back-propagation algorithm \cite{mitchell1997machine} in standard situations and are deactivated when an emergency happens; hence evacuees will be directed to exits over their normal walking paths; however, back-propagation algorithms suffer from slow learning rate and easily converging to local minimum; furthermore, this model cannot react to the spreading of a hazard. The study in \cite{li2010multiobjective} employs a genetic algorithm \cite{john1992holland} to minimise the total evacuation time, travel distance and number of congestion encountered during an evacuation process; non-domination sorting \cite{deb2000fast} is used as no priori knowledge is available to determine the weight of the three goals; the initial ``chromosomes'' are paths found by the $k$-th shortest path algorithm \cite{eppstein1998finding} and are incrementally evolved to feasible solutions through crossover and mutation with respect to the path length, congestion level and hazard intensity; as an evolutionary approach, this algorithm has advantages in solving multi-objective optimization problem (MOP) \cite{saadatseresht2009evacuation}; however, the computational overhead is relatively high due to the path-finding and the evolution process. The research in \cite{pan2005multi} adopts a variation of particle swarm optimization (PSO) to search routes and adjust velocity during evacuations; occupants are viewed as particles to search exits; once an exit is discovered, all the other particles will move towards it while keep their moving inertia to expand searching space; if more than one exit is found, particles will choose the nearest exit as the destination; nevertheless, use occupants directly to explore paths may cause severe injuries and fatalities; meanwhile, this algorithm may induce seriously congestion and oscillation problems. Inspired by the bee colony foraging behaviour, the work in \cite{samadzadeganbiologically} uses bee colony optimization \cite{karaboga2005idea} to displace evacuees in hazardous areas to safe areas during an emergency evacuation; hives, food sources and bees represent safe areas, hazardous areas and evacuees, respectively; evacuees select a safe area with regard to ``attractiveness'' which is determined by the distance to the area and the distribution of people in hazardous areas; once an evacuee determines a target, it will recruit other evacuees by sharing information of the devoted area; this algorithm obtains a robust evacuation plan at the expense of relatively high communication overhead.

Since many of the current emergency response systems are based on wireless sensor networks, routing protocols that are initially used for packet networks have been borrowed or adapted to direct evacuees and improve communication quality in hazardous environments. For instance, the research in \cite{filippoupolitis2010emergency} presents an emergency support system built on top of a WSN to guide evacuees out of a confined space in a real time fashion; the embedded emergency navigation algorithm is inspired by the Cognitive Packet Network routing protocol \cite{gelenbe2001towards,gelenbe2001design}, which was initially designed for large-scale packet networks; different from the original CPN that contains three types of packets: smart packets (SPs), dumb packets (DPs) and acknowledgements (ACKs), the variant only consists of SPs and ACKs; SPs are sent from each sensor nodes in the WSN to search egress paths and collect hazard information in a distribute manner with their predefined goals; when a SP reaches an exit, which means an egress path has been discovered, an ACK will be generated and bring back the ID and the hazard information of each sensor node along the path to the source node that emits the SP; when the ACK reaches the source node, it will update the QoS level of the discovered path by using a rolling average mechanism, which sums the newly discovered QoS and previously stored value in a weighted manner; in attempting to efficiently find the route with the best quality of service (QoS), when a SP arrives at a sensor node, it will decide its next hop by m-Sensible routing algorithm \cite{gelenbe2003sensible}; a QoS metric is defined as sensitive if its value is affected by the traffic through the path, such as congestion level; on the other hand, a QoS metric is insensitive if its value is independent on the traffic assigned to the path, such as path length or number of hops on a path; m-Sensible routing algorithm calculates the probabilistic choices of all the neighbour nodes based on the QoS information brought back by the previous SPs; hence, future SPs decide their next hop by yielding the probabilistic choices of the neighbour nodes obtained by the m-sensible policy; the probability of choosing a neighbour node is determined by $\frac{E[L(n_{\pi(u)},n_{\pi(j)},n_{\pi(e)},\pi)]^{-m}}{\sum_{i = 1}^{N_n}E[L(n_{\pi(u)},n_{\pi(i)},n_{\pi(e)},\pi)]^{-m}}$, where $L(n_{\pi(u)},n_{\pi(j)},n_{\pi(e)},\pi)$ represents the effective length of a path $\pi$ from source node $n_{\pi(u)}$ via a neighbour node $n_{\pi(j)}$ to the exit node $n_{\pi(e)}$; term $E[L(n_{\pi(u)},n_{\pi(j)},n_{\pi(e)},\pi)]$ represents the expectation of $L(n_{\pi(u)},n_{\pi(j)},n_{\pi(e)},\pi)$; term $N_n$ represents the number of neighbour nodes of the node $n_{\pi(u)}$; it is proved in \cite{gelenbe2003sensible} that the QoS increases with the increase of term $m$, hence, an $m+1$-sensible routing policy provides better QoS on the average than the m-sensible policy; to enhance the stabilisation of network, a predefined ``measurement discard threshold'' is set to discard the reported QoS (effective length) that is smaller than the threshold. Since communications which are essential in an evacuation process can easily malfunction due to the hazard, the research in \cite{gelenbe2012wireless} proposes a resilient emergency support system (ESS) to disseminate emergency messages among evacuees and direct evacuees out of a confined space with the aid of opportunistic communications (Oppcomms); the proposed system is composed of pre-deployed sensor nodes (SNs) to collect environmental information and mobile communication nodes (CNs) which are portable devices carried by evacuees; to locate evacuees, each SN contains a location tag and can periodically send a location message (LM) to CNs in proximity; when a SN detects a hazard, an emergency message (EM) will be generated and broadcast to CNs carried by evacuees in vicinity by using the epidemic routing \cite{vahdat2000epidemic}; the EM will be stored in these CNs and forwarded to other CNs in contact by the ``store-carry-forward'' paradigm \cite{pelusi2006opportunistic} during the movement of the evacuees; to guide evacuees, each CN stores the building graph in its local storage and updates the edge costs when receiving an EM; the Dijkstra's shortest path algorithm is triggered to calculate the shortest safest path when the graph is updated; experimental results indicate that the proposed system is robust to network failures during an emergency; since Oppcomms are susceptible to malicious attacks such as flooding or denial of service, an extended study \cite{gorbil2012resilience} proposes a defence mechanism that uses a combination of identity-based signatures (IBS) and content-based message verification to detect malicious nodes. However, network routing protocol based algorithms normally make decisions based on the collected sensory information rather than the predicted situation of a path. Therefore, when evacuees traverse that path, the situation could have changed owing to the highly dynamic nature of an evacuation process, which normally induces a delayed feedback loop between living sensory data and routing decisions. Similar to \cite{filippoupolitis2010emergency}, the research in \cite{BiDesmetGelenbeISCIS2013} also borrows the concept of the cognitive packet network to calculate evacuation paths for evacuees with the aid of an on-site WSN in a distributed manner; however, rather than using m-Sensible routing algorithm, the random neural network (RNN) algorithm \cite{gelenbe1993learning} is utilised as the decision-making algorithm for the SPs to explore the environment; each sensor node in the WSN is considered as a CPN node, in which a RNN is deployed to direct the passing-by SPs and a mailbox is used to store the discovered paths and the associated QoS measurements; the RNN consists of neurons that are associated with each potential forwarding direction of SPs; each neuron possesses an excitation probability to indicate the quality of the forwarding direction, and the neuron with the highest excitation probability corresponds to the optimal forwarding direction; when an evacuation process begins, CPN nodes continuously sends out SPs or relays SPs from other CPN nodes; when a SP reaches a CPN node, it can either select the forwarding direction corresponding to the most excited neuron or drift randomly to search new routes; as a SP arrives at an exit, an ACK will be generated to backtrack the discovered route in a loop-free manner; when an ACK reaches a CPN node, the training process of the local RNN will be triggered and the excitation probability of each neuron will be updated based on the learning mechanism of the RNN; the discovered routes will be stored in the local mailbox and sorted by quality; evacuees in the vicinity of a senor node always will be transferred the top-ranked route as their evacuation route; since each SP can gain ``experience'' from previous SPs, the CPN can rapidly discover the optimal or near-optimal evacuation routes by emitting very few packets \cite{Desmet2}. The studies in \cite{biandgelenberouting2014,birouting2014} extends the work in \cite{BiDesmetGelenbeISCIS2013} and makes use of the feature of CPN to develop a multi-path routing algorithm for different categories of evacuees (prime-aged people, aged people, children or ill people, and disabled people in electric powered wheelchairs) with respect to their specific requirements; since each SP can search a distinct path based on its pre-defined goal function, during the evacuation process, various types of SPs are sent out to search distance-orient paths, time-orient paths, safety-oriented paths and energy-efficiency oriented paths for the associated evacuees. On top of the work in \cite{birouting2014}, The work in \cite{akinwandebigelenbedynamic2015} designs a cooperative strategy that divides evacuees into health-oriented evacuees and evacuation-time-oriented evacuees, and can adjust the routing strategy of evacuees when their ``virtual health value'' fulfills a certain condition; the use of the strategy is proven to be more sensitive and adaptive to sudden changes in the hazard environment such as abrupt congestion or injury of civilians.

By inferring the spreading rate and direction of the hazards, prediction-based algorithms predict the future status in the hazardous areas and reduce the fatality rate by avoid evacuees from traversing paths with high potential risk level. For instance, the research in \cite{hasofer2001stochastic} proposes a Monte-Carlo stochastic model to predict the spreading of the fire hazard and the movement of evacuees during an evacuation process; the targeted building is represented by two graph models composed of nodes (compartments) and edges (passageways), one for fire spread modelling and the other for occupant egress modelling; a discrete hazard function based on Bernoulli trials is used to mimic the propagation of the fire; a Bernoulli trial, which is a random experiment with two possible results: ``success'' and ``failure'', and the probability of success or failure is constant whenever the experiment is conducted, is performed at each time step to mimic the transmission of fire from a compartment to another; each edge of the two graph models is associated with a ``defective'' random variable to represent the time cost for evacuees or the fire hazard to traverse this edge; these defective random variables, which take the value ``infinity'' with non-zero probability, are used to simulate the phenomena such as evacuees cannot reach the next node owing to the capacity limitation or fire cannot reach the next node due to fire fighting activities. The study in \cite{barnes2007emergency} proposes a WSN based distributed navigation algorithm to search safest routes for evacuees by maximising the time an evacuee will remain ahead of the hazard while traversing the route; each sensor will maintain two weighted graphs of the built environment, a ``hazard graph'' and a ``navigation graph''; in the hazard graph, nodes represent the locations of sensor while edges represent the possible movement directions of the hazard (for example, hazard may spread through walls or along corridors); the weight of an edge is the shortest time for the hazard to propagate along the edge, these information can be obtained from off-line hazard simulations \cite{olenick2003updated} or estimated by emergency engineers; in the navigation graph, nodes represent the sensor locations while edges represent the possible movement directions of evacuees; the weight of an edge is the longest time for an evacuee to traverse the edge; when a fire breaks out, sensors that detect the hazard will broadcast the fire source location over the senor network, then each sensor calculates the safest path to exit by maximising the overall difference in time between an evacuee arriving each node on a path and the hazard reaching these nodes. Since it is difficult for fire-fighters to be aware of the actual conditions in a built environment during a fire disaster, the research in \cite{han2010firegrid} presents a novel e-infrastructure to infer the spreading of hazard based on predictive models and living sensory data in a faster-than-real time manner; the system consists of on-site sensors including smoke detectors and temperature sensors, and off-site computational models that are deployed on High Performance Computing (HPC) resources; gathered sensory data is used as inputs into a Monte-Carlo fashion fire spread model called K-CRISP \cite{koo2008sensor} to prediction the movement of fire and smoke; the results are interpreted by using a knowledge-based reasoning scheme within an agent-based command-and-control layer; the outputs are transmitted to fire-fighters for reference. To deal with the highly uncertainties during an evacuation process in an unfamiliar built environment, the work in \cite{radianti2015spatio} proposes a Dynamic Bayesian network (DBN) \cite{murphy2002dynamic} based spatio-temporal probabilistic model to capture the uncertain nature of the hazard and crowd dynamics, and forecast the movement of evacuees; the integrated hazard and crowd evacuation DBN model is composed of a hazard model, a risk model, a behaviour model, a flow model and a crowd model; each model embeds a DBN and is subject to the Markov condition; by using the integrated hazard and crowd evacuation DBN model, the relations between the location of evacuees and the hazard status of each location (dormant, growing, developed, decaying and burnt-out) are tracked and predicted over adjacent time steps; hence, the probabilistic risk level of each location can be derived from the model; the egress paths are calculated by Dijkstra's algorithm with respect to the estimated risk level of each location with the purpose of minimising the overall fatality rate. These algorithms are developed in recent years and are quite promising owing to the increasing popularity and tremendous computing power of the cloud computing paradigm. Since the performance of many emergency navigation algorithms is sensitive to various initial conditions (e.g. the initial distribution of evacuees, congestion level, type of disaster and initial disaster location) and the choice of certain parameters, the research in \cite{BIGELENBEIEEEPERCOM2015} presents a faster-than-real-time simulation based routing algorithm to predict the future situation before guiding evacuees to exits when a disaster breaks out; instead of guiding evacuees in a real time manner, this algorithm borrows the tremendous computational power of the cloud based simulator to rapidly identify the potential death victims by predicting the future movements of evacuees and spreading of the hazard; then, an iterative based algorithm is employed to gradually search appropriate paths for these potential death victims; finally, all the calculated paths are sent to evacuees for instruction and evacuees could follow the paths in source routed manner (do not need to switch paths during the evacuation process).

\section{Emergency Search and Rescue Planning}
\label{emergencysearchandrescueplanning}

Originating from maritime search and rescue operations, search and rescue planning in emergency situations has motivated considerable research over the last several decades owing to the unfortunate increasing threat of both manmade and natural disasters. During a disaster-related emergency evacuation, evacuees may become immobilised and incapacitated due to injuries or obstacle contact. Therefore, to reduce the fatalities, various emergency management systems have been proposed to detect the location of incapacitated evacuees and dispatch rescuers to perform rescue operations. The main challenges of a rescue operation are threefold. The first challenge is how to efficiently search and locate injured evacuees or other objects in unknown environments, especially on how to coordinate the activities among various rescuers. The second is to design an appropriate rescuer assignment algorithm to allocate rescuers to injured evacuees in a real time and computationally efficient fashion under the highly dynamic hazardous environment. This is actually a NP-hard assignment problem \cite{gelenbe2014rescuerallocation}, which aims to minimise the overall potential cost for the rescue operation. The third challenge is to search desired paths for rescuers and victims to fulfil their specific requirements, which is difficult since the ``quality'' of a path is affected by the spreading of the hazard, the dynamic congestion level, the movements and behaviours of evacuees and other rescuers on the path. In the following subsections, we will summarise various systems and algorithms that have been proposed to meet the aforementioned challenges.

\begin{figure}[!ht]
\centering
\includegraphics[width=0.6\textwidth]{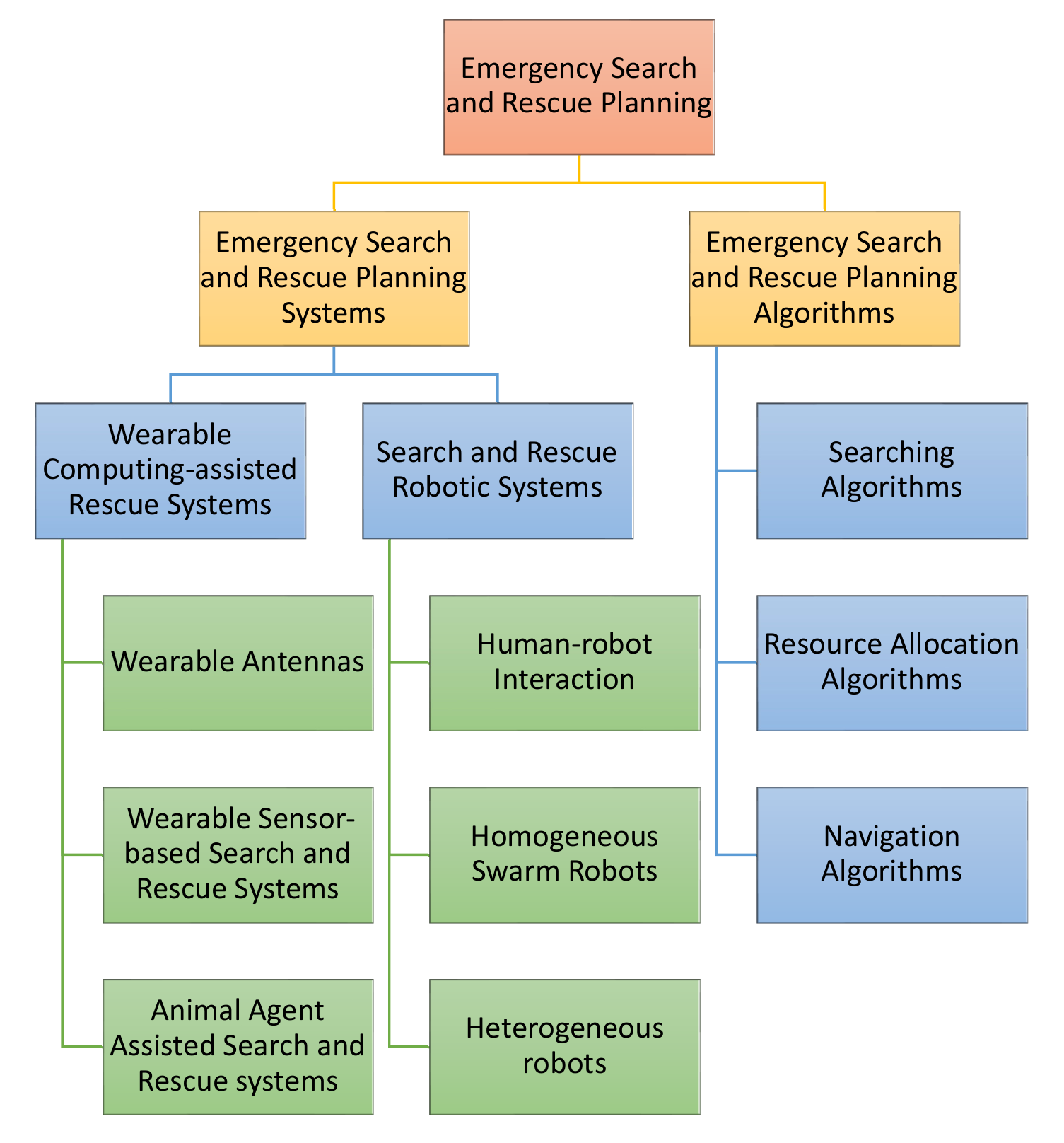}
\caption{A tree diagram explaining the structure of Section \ref{emergencysearchandrescueplanning}.}
\label{fig: EmergencySearchandRescuePlanning}
\end{figure}

\subsection{Emergency Search and Rescue Planning Systems}
\label{emergencysearchandrescueplanning:systems}

Based on the facilities used, emergency search and rescue planning systems can be divided into two categories, wearable computing-assisted rescue systems and search \& rescue robotic systems. Wearable computing-assisted rescue systems concentrate on providing various enhancements for emergency personnel (e.g. firefighters, rescuers) with the aid of wearable devices to increase the efficiency of rescue operations and improve the safety of emergency responders. On the other hand, disasters that create harsh environments with extreme temperature, toxic substances or various obstacles have exposed the unsafety and inefficiency of the human-centered search and rescue planning systems. These limitations have therefore inspired the development of the robot-centered search and rescue planning systems, which employ various mobile robots to conduct rescue operations.

\subsubsection{Wearable Computing-assisted Rescue Systems}

With the rising interest in body-centric wireless communications, which has also been standardised as a part of the fourth generation mobile communication systems (4G) \cite{rais2009review}, considerable research has been dedicated to develop low cost, light-weight wearable antennas to maintain and improve emergency communications. The work in \cite{orefice2016electrically} implements an electrically-small wearable antenna, which is integrated into the sleeve of a jacket, to monitor the positions of emergency rescuers; the designed frequency is at around 860 $MHz$ since higher frequencies could suffer from shadowing caused by obstacles; the antenna consists of an electromagnetically coupled square patch and a central shorting pin; to reduce the cost and weight of the antenna, various conductive textiles such as mixed cotton-steel threads are employed and tested as the material for both the patches and the ground plane. The research in \cite{lilja2013body} designs a wearable antenna integrated with the existing Cospas-Sarsat, a satellite-based search and rescue system that provides distress alert detection and information distribution services by locating and communicating with activated personal locator beacons, to provide emergency alert and location information for maritime rescue teams. The antenna is embedded into a life vest, and the moisture-absorption characteristics of the textile can affect the anntenna's performance. The textile material used must be evaluated for ``Moisture regain'', i.e. their moisture-absorption rate, measured by the relative weight increase when kept in a high-humidity environment, and the antenna is placed on a foam substrate with low moisture-absorption rate, a thin carrier foil with an inkjet-printed antenna pattern, and a cover fabric to protect the antenna from wear and tear, and from water infiltration.

The study in \cite{li2009ern} presents an emergency rescue navigation system to direct firemen to ``key corridors'' to eliminate fire and congestion caused by evacuees or obstacles generated from hazard. The rescue navigation system is composed of a remote control center to generate instructions for firefighters, an on-site WSN to provide hazard \& location information, as well as 802.15.4 compatible personal digital assistants (PDAs) carried by firefighters to communicate with the control center and the WSN via naive flooding or the opportunistic flooding strategy. To determine the ``key corridors'' that affects the efficiency of the evacuation most, this problem is firstly converted to a ``maximum flow problem'' by connecting each normal node (represent the location of a sensor) with a virtual source and each exit with a virtual sink; the edges that can maximise the amount of flow passing from the source to the sink are then determined by the \emph{Max-flow Min-cut theorem}; finally, a breadth-first-search strategy is used to search the key corridors with existing hazard or obstacles from the afore-determined edges. The research in \cite{huang2005CenWits} presents a wearable sensor-based search and rescue system, namely ``CenWits'', to locate lost or injured hikers in wilderness areas with an opportunistic relaying scheme; each hiker carries a wearable sensor with a built-in GPS receiver that can provide position information and an integrated RF transceiver to exchange ``witness information`` (movement and location information) with other hikers in vicinity; to maintain the communication between hikers and the rescue center, access points (APs) are deployed at locations of interest that hikers are likely to pass through, and convey the witness information stored in the wearable sensors to the rescue center; due to the limited battery power and memory of wearable sensors, CenWits also provides an adaptive data storage strategy to optimise the trade-off between battery power and memory utilisation with the aid of a dynamic grouping and partitioning mechanism; if the remaining battery power of wearable sensors is low, hikers are partitioned into groups where only the group leaders store all witness information and communicate with the APs while the rest of group members are set to sleep mode; if the remaining battery power of wearable sensors is sufficient while the remaining memory is running low, hiker groups are further divided into sub-groups where each group member stores a subset of the witness information.

As an effort to decrease the likelihood of emergency events from the ``prevention'' aspect, the research in \cite{wu2008waiter} presents a wearable personal healthcare and emergency aid system, namely ``WAITER'', to monitor the health status of users with wearable vital signal sensors and alert the remote healthcare center when an emergency occurs; aiming at substituting the labour-intensive caregiver aid, this system comprises of a body-worn vital signal sensor to collect health status data, a mobile phone to perform on-site computation and storage operations for the raw data and a remote healthcare center to provide timely medical aid; the vital signal sensor is integrated into a Bluetooth ear-set which consists of a heart beat sensor, a motion sensor, a body temperature sensor and a Bluetooth wireless communication device; since the high energy consumption of the wireless data transmission, raw data gathered by the vital signal sensor is first transmitted to the mobile phone for refining and validation via the Bluetooth transceiver; once an emergency is detected, the mobile phone generates an alert with the filtered data and then sends to the healthcare center via its GSM module.

Owing to the superior performance of animal agents during search and rescue operations in terms of mobility, energy utilisation efficiency, sensory acuity and intrinsic cognitive capacity, much effort has been dedicated to animal agent assisted search and rescue systems. The research in \cite{bozkurt2014toward} employs cyber-enhanced working dogs to locate and reach survivors trapped under rubble in the aftermath of large-scale disasters; this system contains three components: a smart harness worn by a working dog to monitor the surrounding environment and the canine, a remote computer carried by the handler to analyse and control the behaviours of the canine, and mobile base stations such as unmanned vehicles to maintain the communication between the working dog and the handler; the smart harness is equipped with various sensors and actuators, including DC vibration motors to receive haptic commands, a mini-speaker to issue aural commands, a treat dispenser to reward the canine for desired actions, accelerometers and gyroscopes to monitor the posture of the working dog, as well as a GPS receiver, microphones and cameras to gather environmental information; on the other hand, the remote computer is employed to identify the posture and behaviours of the canine remotely, the activities of a canine are classified into 5 static postures (sitting, standing, lying down, standing on two legs, and eating off the ground) and 3 dynamic behaviours (walking, climbing stairs, and walking down ramps), three hidden Markov models, which are each associated with one of the dynamic behaviours, are utilised to classify the dynamic behaviours of a working dog; the starting state probabilities and transition probabilities of the hidden Markov models are estimated with the iterative Baum-Welch algorithm; the input sensory data that cannot be categorised into the 3 dynamic behaviours will be refined with a moving average filter and then identified as a static posture. In attempting to penetrate the voids and narrow gaps under the collapsed ruins after an major earthquake, the study in \cite{bozkurt2016biobotic} employs cockroaches to search trapped survivors and map the under-rubble environment; to instruct the locomotory behaviours of a cockroach, fine wire electrodes are implanted into the antennae of the cockroach to perform neurostimulation-based locomotion control; the searching strategy of this system is to keep each cockroach moving and exploring naturally within a defined area; when the cockroach reaches the boundary of the area, left-turn or right-turn commands will be sent from in the form of neurostimulation pulses via the fine wire electrodes; each cockroach is also mounted with three directional microphones to locate the survivors by using the ``received signal strength indicator'' and the ultrasonic ranging technologies; to construct a robustness map of the environment, a novel topological mapping approach is used to first extract topological features from encounters among the cockroaches, and then refine the persistent features from the generated encounter map.

\subsubsection{Search and Rescue Robotic Systems}
\label{SearchandRescueRoboticSystems}

Since the human access to the victims in the aftermath of a disaster such as an earthquake or radioactive leakage can be time consuming and may induce further casualties \cite{satoshi2009earthquake}, the use of robots that can be released rapidly to locate and rescue victims has drawn considerable attention in recent decades, and has gradually become a research domain after the Great Hanshin earthquake and the Oklahoma City bombing in 1995 \cite{davids2002urban}. Nowadays, research on ``search and rescue robotic systems'' has evolved as a major branch of ``search and rescue'', especially for urban scale structural collapse environments \cite{barbera1996urban}. According to the different operating environments, search and rescue robots can be divided into autonomous underwater vehicles (AUVs), and urban search and rescue (USAR) vehicles \cite{stopforth2008survey}, which contain various types of unmanned ground vehicles (UGVs) and unmanned aerial vehicles (UAVs). This literature review focuses on the research of USAR robotic systems, and the research on AUVs is excluded due to its distinctive emphases caused by the turbid underwater environment.

Due to the limitations on the mobility (tracks, wheels or combination of both) and intelligence, most of the current search and rescue robot systems are not advanced enough for fully autonomous operations. Hence, a research direction is to investigate the human-robot interaction during search and rescue operations. The first known use of USAR robots for an actual unstaged rescue mission occurs during the World Trade Center collapse \cite{casper2003human}, when various models of tele-operated robots from industry, military and academia were invited to assist the rescue operation; these robots carried various sensors such as cameras, microphones and speakers to provide real time video and audio for remote human operators for victim identification, and various actuators such as a robot arm with a gripper attached topside for victim extrication, a medical tube for providing fresh air or water; moreover, this work conducts a post-hoc analysis on the performance of the human-robot interaction during the rescue mission based on the video tape and field notes; the analysis reveals the pressing needs of improving the mobility, intelligence and assistive interfaces of rescue robots, as well as reducing the number of people needed to operate a robot since the performance of operators are significantly affected by cognitive fatigue caused by lack of sleep. The research in \cite{barzin2014learning} presents a semi-autonomous robot controller to share and co-execute tasks with remote human operators during the exploration of unknown disaster scenes; instead of using fully human-supervised robotic control which may induce cognitive and physical fatigue, disorientation and incorrect judgements, a MAXQ hierarchical reinforcement learning (HRL) based controller is developed in this study to determine whether the robot or the human operator should conduct a certain rescue task with respect to the prior knowledge and experiences; the MAXQ algorithm decomposes the overall SAR tasks into four types of subtasks that can be learned simultaneously: local navigation (obstacle avoidance), navigate to unvisited regions, victim identification, and human control; each subtask can be further decomposed into child tasks recursively until each child task is a primitive action of the robot; then a task hierarchy can be built by iteratively mapping all the possible states of a subtask to its child tasks; each state-action pair is associated with a action-value function, which represents the expected cumulative reward for performing the action in the given state when following the task hierarchy; the MAXQ HRL-based learning algorithm updates the values of each state-action pair based on the received positive or negative rewards (e.g. a positive reward is given to the local navigation subtask when a collision is avoided) and the action with the highest value is selected as the optimal policy.

One of the major challenges that teleoperated or semi-autonomous rescue robots face today is the high human-to-robot ratio (The human-to-robot ratio denotes the number of people needed to operate a robot.), which reaches $2 : 1$ or even $3 : 1$ in the current state of practice \cite{murphy2011100}. The high human-to-robot ratio restricts the scale of rescue robots to be deployed in the devastated region since too many operators can significantly increase the logistic burden and the training cost. One feasible solution to this problem is the use of homogeneous swarm robots, which are easier to control and understand. For instance, the research in \cite{halasz2007dynamic} employs a team of homogeneous robots to search and explore several disaster sites; because of the possible environmental changes and robot failures in disaster sites, the robots are required to continuously redistribute to fulfill the desired population fraction of robots at each site; furthermore, due to the harsh physical conditions and resource limitations, the inter-robot communications can be unreliable and costly, and are not easily applicable during an search and rescue operation; hence, this study proposes a task allocation policy to redistribute the robots from an initial distribution to the desired distribution without inter-robot communications; the relation between population fraction of robots at each site and the elapsed time from the start of the process is expressed by a bunch of ordinary differential equations, under the assumption that the graph model of the devastated region and all the transition probabilities for a certain robot to traverse from a certain site to another are known to all the robots; by solving the ordinary differential equations, each robot can compute the time point when the system reaches the  desired distribution in a decentralised manner. Inspired by the self-organised behavior of social insects, the study in \cite{berman2009optimized} extends the work in \cite{halasz2007dynamic} and proposes a stochastic task allocation strategy to assign a swarm of homogeneous robots to various tasks during an search and rescue operation without inter-robot communications; the latency (transition time) for a robot to switch from one task to another is modeled by an Erlang distribution to capture the fact that the transition times have positive, minimum possible values and a small likelihood of being large because of low battery power, accidents or breakdowns; hence, the behaviours of the swarm are modelled by a bunch of delay differential equations with population fractions as variables; these delay differential equations are then converted into equivalent ordinary differential equations by fitting the transition times with empirical values; by solving these ordinary differential equations, each robot can calculate the time point when the swarm reaches the  desired population fractions at tasks in a distributed manner; to ensure the fast convergence of the system, the transition probabilities for a certain robot to switch from a certain task to another are optimised with various optimisation tools.

Owing to the diverse and complex tasks in rescue operations, most of the recently proposed USAR robot systems consist of heterogeneous robots with different capabilities instead of homogeneous robots. For instance, the research in \cite{iwano2004rescuerobot} proposes a robot group which consists of three types of robots to rescue victims during nuclear plant accidents; information-collecting robots are used to gather the location and posture information of victims; traction robots are employed to manipulate the victims' posture to an easy-to-carry state; transportation robots are coupled to form a stretcher to carry victims out of the hazardous area; to manipulate the posture of a victim to a desired attitude, the body of the victim is simplified as a link model where limbs are considered as links and joints are considered as vertices; the angle deviation between the targeted posture and the initial posture is calculated by the dot product of the vectors associated with these two postures; a robotic hand is attached on the traction robots to grasp and move the limbs of the victim; to avoid the human body from colliding with the walls during the transport process, a Generalised Voronoi Graph based motion path planning algorithm is utilised to compute a path that is equidistant from the surrounding walls and the obstacles. The study in \cite{ventura2012search} presents a USAR robot system which contains aerial robots and a land robot to detect the potential survivors in high-destruction locations after an earthquake; the aerial robots are responsible for mapping the interested area, searching desired paths with less debris and more victims for the land robot, and maintaining the communications between the land robot and remote human operators; on the other hand, the land robot is used to detect survivors with specific sensors such as onboard cameras; to climb over debris or partially destroyed stairs, the land robot, which is composed of a main body and a frontal body with two side tracked wheels, is designed with the shape-shifting ability; the relative angular orientation between the main body and the frontal body is adaptable to adjust the mass center of the robot and provide extra traction power; moreover, to free the human operators from tedious and difficult tasks, this system also provide an immersive 3D head-mounted display to improve the situation awareness of the human operators, and a visual servoing based automatic docking subsystem to dock and undock the robot to the tether, which carries an electrical power supply and communication devices.

In addition, to verify the effectiveness of the proposed robotic systems, various real and simulated scenarios have been designed in the last two decades. The research in \cite{jacoff2003test} designs a test arena with collapsed structures to evaluate the agility, planning or mapping algorithms and sensing ability of rescue robots. To mimic large scale disasters, a proving ground namely ``Disaster City'' with a size of $210,000 m^2$ has been built by the US Federal Emergency Agency (FEMA) to conduct rescue drills for robots \cite{ventura2012future}. The European counterparts are the Rescue Robot Fieldtest which simulates a traffic accident involving a hazardous material truck and European Land Robot Trials (ELROB) that can imitate a large scale incident such as a terrorist attack at a public event. The RoboCupRescue simulator in \cite{tadokoro2000robocup} integrates various disaster models and intelligent agents to build a virtual environment, and a real-world interface is designed to connect the virtual world with facilities in the real disaster field; evacuees' decisions and rescue plans are transmitted back to the simulator via a human-machine interface.

\subsection{Emergency Search and Rescue Planning Algorithms}
\label{emergencysearchandrescueplanning:algorithms}

The state-of-the-art emergency search and rescue planning studies are originated and derived from research in maritime search and rescue activities, which has been a significant research topic since 1970s. Related algorithms in emergency search and rescue planning are commonly under the assumption that multiple agents (rescuers, robots, etc.) are involved. The reason behind this is twofold. First, time efficiency is the dominant factor for the success of a search and rescue operation, and multi-agent based search and rescue operations are obviously more efficient than single-agent based search and rescue operations. Second, multi-agent systems are generally more cost-efficient and feasible than a single agent with all the capabilities \cite{gautam2012review}. In recent decades, related algorithms have aroused a new interest in not only searching victims in hazardous environments, but also de-mining \cite{gelenbe1998autonomous} and planetary exploration \cite{landis2004robots}. The philosophy behind team-based search and rescue is to convert a complex problem into simpler sub-problems that are more efficient to solve. These algorithms are be classified into three types: first, searching algorithms, which are dedicated to search and locate injured evacuees in unknown environments, specifically concentrating on coordinating the search hehaviours of rescuers to produce ``swarm intelligence''; second, resource allocation algorithms, which aim to assign rescuers to injured evacuees in a desired manner when the locations of evacuees are revealed; third, navigation algorithms, which focus on discovering appropriate paths for rescuers when the locations of evacuees are known. In this section, we review the related algorithms of first two categories. The third category is excluded here since all the on-line emergency navigation algorithms are applicable to this case, and the detailed review can be found in Section \ref{emergencynavigationalgorithms: onlinealgorithms}.

\subsubsection{Searching Algorithms in Emergency Search and Rescue Planning}

The research in \cite{pugh2007multirobotpso} proposes a particle swarm optimisation (PSO) based multi-robot system to find targets by gradually optimising the pre-defined goal function; each robot is modeled as a particle whose velocity and position is determined by the neighbouring particles and the previous best position of this particle. Similarly, the study in \cite{zhu2011pso} designs a PSO based multi-robot system to search targets and avoid obstacles in an unknown environment; specially, a relative coordinate system is used to avoid the dependence of the precise global location of robots. By employing a bio-inspired random search behaviour called l{\'e}vy flight, the research in \cite{sutantyo2010multi} presents an efficient multi-robot searching algorithm to search targets in an unknown environment; an artificial potential field is used to generate repulsive forces to disperse robots amongst the area of interest. The research in \cite{baranzadeh2015distributed} deploys robots in a triangular grid pattern, which can minimise the number of robot required for the coverage of an area, to search an unknown zone; the targeted area is firstly partitioned into equilateral triangles and then the robots will search along the edges of the triangular grid. To minimise the total travel distance during the search process,  the study in \cite{cavalcante2012local} utilises an auction based algorithm to assign desired tours to robots. The research in \cite{marjovi2009multi} presents a multi-robot coordination algorithm to identify fire sources in an indoor environment; exploration frontiers are extracted and assigned to robots to minimise the overall exploration time. Frontier-based searching algorithms such as \cite{burgard2005coordinated,barzin2014learning} normally concentrate on assigning robots to the boundary line between the visited and unvisited areas; the search direction is commonly affected by the cost for a robot to explore a boundary area (e.g. the probability of an obstacle or another robot being present) and the cost for the robot to reach the target boundary area (e.g. the distance from the target boundary area). The study in \cite{gelenbe2010search} investigates an interesting research issue of searching a fixed object in an unknown environment with a number of searchers performing Brownian motions; the searchers are sent out consecutively from the source to locate the object at a finite distance $D$, the search space is unbounded and the distance $D$ is unknown; each searcher is subject to a finite life span and could be destroyed or disabled during the search; a time-out is set by the source to eliminate searchers that search for a long time without discovering the object; when the time-out elapses, a new searcher will be sent out to replace the ``lost'' searchers; a closed form expression for the average search time is derived as a function of the distance between the source and the object (``$D$'') by modelling the search process as a multidimensional Brownian process; the experimental results show that the average search time is affected by number, life span, routing uncertainty, destruction rate of the searchers. The research in \cite{lau2007immuno} presents a robotic search and rescue system to search for victims in rubbles with lost cost homogeneous robots carrying thermal array sensors; a general suppression control framework, which is inspired by the suppression mechanism of the human immune system, is proposed to regulate the searching and exception handling behaviours of the robots.

\subsubsection{Resource Allocation Algorithms in Emergency Search and Rescue Planning}

The study in \cite{gelenbe2014rescuerallocation} presents a resource allocation algorithm on the basis of the Random Neural Network (RNN) with synchronised interactions \cite{RNN2008} to assign rescuers to trapped victims in a fire-affected building; to optimise this task assignment process, several effect factors such as the cost of assigning a rescuer to a victim (e.g. the distance between the rescuer and the victim), the probability that a rescuer fails to save a victim, and the associated penalty that the rescuer fails to rescue the victim are taken into consideration, and are transformed into a goal function which reflects the trade-off between the total successful rescue costs and the failure penalties; the RNN is utilised as a fast optimisation algorithm to minimise the goal function with a gradient descent learning procedure; in the RNN, each possible rescuer-victim pair is considered as a neuron, and the neuron with the highest excitation probability after the training process is selected as the decision. The research in \cite{su2015dynamic} proposes a task allocation algorithm for heterogeneous agents in hazard environments with respect to time, space and communication restrictions; to reduce the communication load and links among agents, each agent elects the agent with the most direct neighbours within its communication range as the communication network leader; the elected network leader will be responsible for maintaining the communication among intra-network agents and other network leaders; to ensure the appropriate distance among task locations and the agents, each network leader further divides its coverage area into sub-areas by using the mean shift clustering algorithm; the window radius parameter $h$ in the mean shift clustering algorithm is set to the communication range of the network leader to guarantee that the agents in the same sub-area can always communicate with each other; heterogeneous agents are then allocated into these sub-areas in terms of their abilities and the requirements of tasks within the sub-areas: firstly, the ``similarity'' value of each agent sub-area pair is calculated by the dot product \cite{kang2012sphere} of the vector of the requirements of tasks and the normalised vector of ability of the agent; secondly, the agent will be assigned to the sub-area with the highest similarity value. As aforementioned in subsection \ref{SearchandRescueRoboticSystems}, the studies in \cite{halasz2007dynamic,berman2009optimized} employ an ordinary differential equation model to redistribute rescue robots among various tasks in a decentralised manner without inter-robot communications; the relation between population fractions at tasks and the elapsed time since the start of the process is expressed by a set of ordinary differential equations; by substituting the desired population fraction in each task into the proposed model, each robot can independently calculate the time point when the system reaches the desired distribution.

\section{Emerging Challenges and Opportunities}
\label{challengesandopportunities}

\begin{figure}[!ht]
\centering
\includegraphics[width=0.6\textwidth]{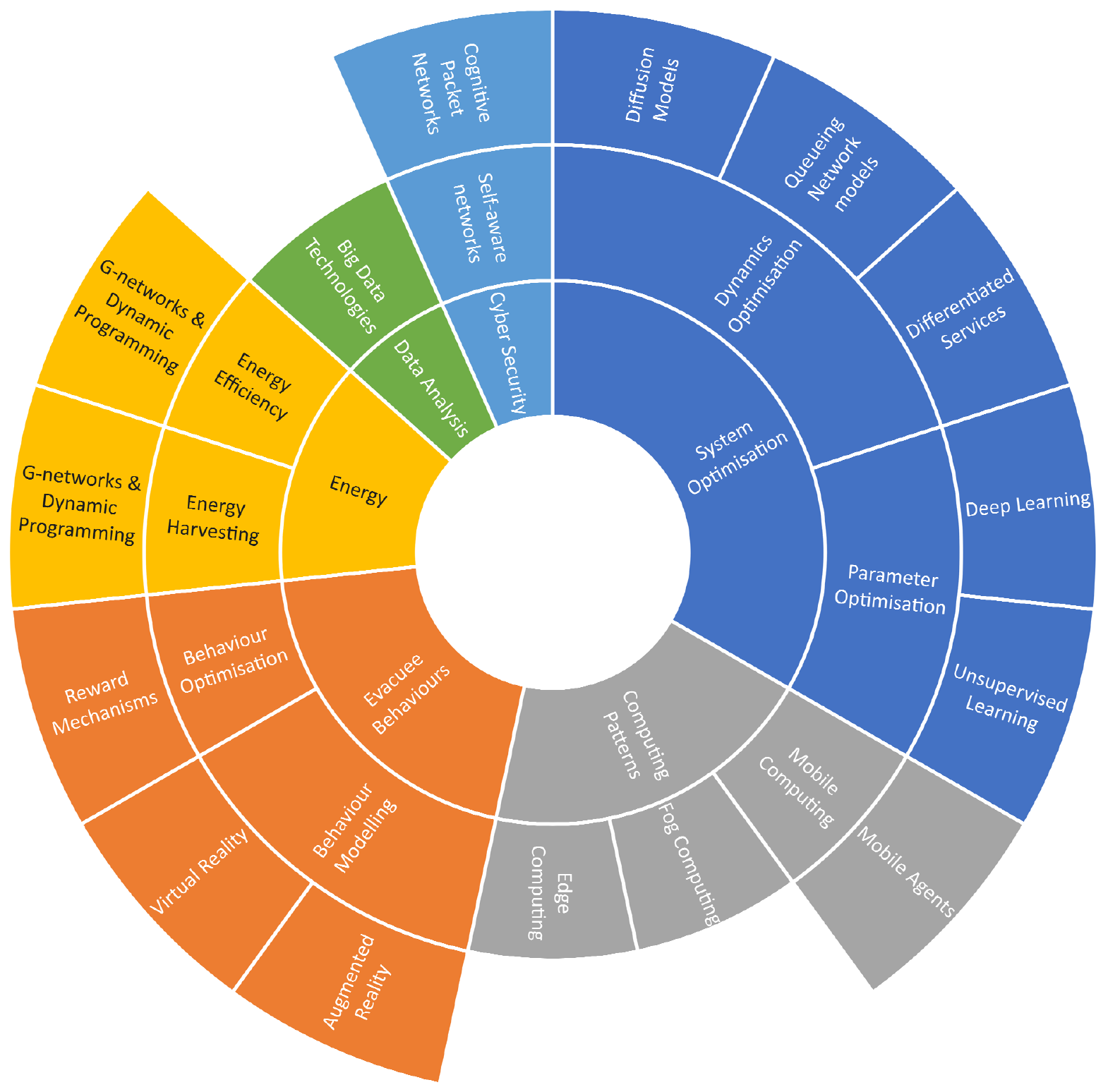}
\caption{A sunburst chart illustrating the potential research directions.}
\label{fig: sunburst}
\end{figure}

After decades of study and exploration, emergency management has become a mature research field. However, due to its open and inclusive nature, new technologies always tend to influence, change or even revolutionise this research area. In this section we discuss open issues and provide possible directions for future work. We also visualise these research directions by using a sunburst chart as shown in Figure \ref{fig: sunburst}.

The emergence of heuristic based algorithms has offered near-optimal solutions for emergency management in a computationally and time efficient manner at the trade-off of optimality and completeness. However, the setting of key parameters in heuristics could play a vital role in system performance and the perfect parameter settings for one evacuation scenario may not suit others. Hence, currently, significant parameters involved in heuristics are mostly pre-configured in a supervised-learning fashion rather than using a unsupervised learning paradigm and therefore may not adapt to uncertain or fast-changing environments, and may even need to be calibrated manually in different scenarios. Taking WSN-based algorithms for example, the work in \cite{Desmet2} has shown that these parameters could contribute significantly to the efficiency of route discovery and information collection. For instance, if the time-to-live of packets is too large, the system will be overburdened with packets that are in effect lost. On the other hand, if the live-time constraint is too small, some distant exits may not be discovered. Likewise, the improper packet rate could induce unnecessary energy consumption when appropriate paths have been discovered and the network situation stays unchanged. Hence, future research can be directed to optimising the setting of parameters from a systemic point of view by utilising proper queueing network models or diffusion models. Potential differentiated services (DiffServ) mechanisms could be also developed to optimise the packet behaviours and satisfy the QoS requirements of different categories of cognitive packets which are associated with diverse classes of evacuees. Furthermore, parameter optimisation algorithms or even parameter free algorithms can be developed by gaining experience from iterations of various simulated scenarios with the aid of newly-proposed machine learning algorithms such as deep learning.

Compared with emergency drills, computer simulations are commonly utilised to investigate the effectiveness of emergency management algorithms due to their time-efficiency, repeatability and cost-efficiency. However, many human behaviours during emergency evacuations have been ignored in the simulations. Hence, future research can also be directed to develop various human-computer interfaces for simulators to generate an artificial hazard environment with virtual reality and augmented reality technologies \cite{gelenbe2005simulating}. By evacuating volunteers from the artificial hazard environment, their emergent behaviours can be collected and empirical collective behaviour models of human beings can be developed from the actual crowd measurements.

One trending in emergency management is to develop artificial intelligence aided algorithms to improve path-finding and resource allocation during standard or emergent evacuations. However, as a typical category of cyber-physical systems, the intelligence of evacuees, as well as the possible pro-social behaviours such as helpfulness and sense of duty, have been excluded in the previous algorithms. As a result, the robustness of these algorithms cannot be ensured due to the high likelihood that evacuees do not follow the instructions. Moreover, unnecessary efforts have been dedicated to the use of AI, while in fact many tasks can be easily accomplished by evacuees in the system via using human intelligence. For instance, the identification of a fire source location could take significant efforts when using AI-based algorithms, which involves in the installation of cameras and the use of computer version related technologies. However, it would take no time or cost for evacuees to identify a fire and report to authorities or the emergency response system by using smart phones. In attempting to incentivise evacuees to conduct cooperative behaviours and avoid destructive behaviours, one future research direction could be dedicated to the investigation of reward mechanisms to integrate the human intelligence of evacuees into the emergency response systems to improve the efficiency, adaptiveness and robustness of these systems. On the other hand,  since most of the previous multi-robot rescue systems only follow simple coordination rules and lack explicit teamwork models or goals \cite{nourbakhsh2005human}, multi-agent technologies can also be utilised to model and develop various cooperative strategies with the aid of queueing network models such as G-networks \cite{G-Nets} or genetic algorithms.

The rapid development of Cloud technologies has facilitated the advancement of cloud-enabled emergency management systems that consist of front-end portable devices and back-end cloud severs. Future research can be directed to further improve the flexibility of the kind of system by leveraging the mobile agent technology, because current cloud-based emergency response systems, which are based on the client-sever paradigm, demand pre-installed services in participating devices \cite{aversa2010cloud}. Moreover, since mobile agents can migrate seamlessly through multiple clouds and different portable devices, a mobile agent-based emergency response system has the potential to reduce communication costs and ease network congestion in large-scale emergency evacuations by dynamically optimising the locations of mobile agents. Moreover, we believe that future emergency management systems should not only make use of the computing power of portable devices and the Cloud, but also other individual devices in the vicinity to offer services with the aid of edge computing and fog computing.

Although the fire alarm system has become a norm for modern buildings, more complicated emergency management systems mostly remain in prototype form and are only verified in simulations or specific test fields. The raising of the concept ``smart city'' has provided a golden opportunity for emergency management systems to be integrated into the smart city ``eco-system'' and take advantage of this tide to achieve leapfrogging. The potential research directions are threefold. First, with the recent increase of manmade disasters, physical attacks are highly likely to be accompanied with cyber attacks. Hence, it is of critical importance to develop self-aware networks to self-defense and maintain communications during an evacuation process. On suggested framework is CPN, which is naturally efficient on defending various forms of malicious attacks such as denial of service attacks \cite{gelenbe2007self}. This is because, unlike the IP protocol, smart packets in the CPN carry the ``full path'' from the source node to the destination node. Hence, CPN can defend denial of service attacks by adaptively dropping attacking packets upstream from the node being attacked via backtracking the full path of attacking flows. Second, due to the energy hungry communication and computation processes inside emergency management systems as well as the fact that wireless sensors are difficult to be replenished, energy efficient and energy harvesting algorithms, such as dynamic programming and G-network models \cite{gelenbe2012energy,gelenbe2014adaptive,gelenbe2014sensor}, can be applied to future emergency management systems to improve the performance and efficiency of these systems. Third, the massive deployment of sensors with the prosperity of the Internet of Things(IoT) and smart city technologies improve the efficient collection of individual data on a vast scale. This suggests a crucial new opportunity to use big data technologies to analyse, model and refine the personal preferences, human collective choices and behaviours,  and general rules with respect to the routinely collected data, and use them as guidance for the future emergency management algorithm and system design.

\section{Conclusion}
\label{conclusions}
In this paper, we provide a systemic review of the emergency management research. In the first section, we review the history and evolution of this field, trace its transformation from a reactive manner to a proactive manner. We also explore the impact of the development of computer technologies, which has shaped the current research. In the second section, we review the emergency search and rescue planning from both system design aspect and algorithm engineering aspect. In the system review part, various wearable computing‐assisted rescue systems, and search \& rescue robotic systems are introduced. In the algorithm review part, representative searching, resource allocation and navigation algorithms are discussed. Similarly, in the third section, we summarise classical emergency navigation systems and algorithms. The systems are classified into human experience driven systems, static WSN based systems, mixed WSN based systems, cloud based systems with WSN, and cloud based systems with mobile phones. The algorithms are divided into off‐line navigation algorithms and on‐line emergency navigation algorithms. The off‐line navigation algorithms are further divided into cellular automata model based algorithms, social force model based algorithms, fluid-dynamics model based algorithms, lattice gas model based algorithms, game theoretic model based algorithms, computer agents based algorithms and animal agent based algorithms. On the other hand, on‐line emergency navigation algorithms are classified into network flow based algorithms, geometric algorithms, queueing model based algorithms, potential-maintenance algorithms, biologically inspired algorithms, routing protocol based algorithms and prediction-based algorithms. In the next section, we discuss the emerging challenges and opportunities under the background of the fast development and prosperity of new technologies such as smart city, artificial intelligence, virtual reality, big data, cyber security and energy efficiency \& harvesting.

\vspace{6pt}




\conflictsofinterest{The authors declare no conflict of interest.}

\bibliographystyle{mdpi}%
\bibliography{bibliography}

\begin{thebibliography}{-------}
\providecommand{\natexlab}[1]{#1}

\bibitem[Drabek and Hoetmer(1991)]{drabek1991emergency}
Drabek, T.E.; Hoetmer, G.J.
\newblock Emergency management: Principles and practice for local government.
\newblock {\em Washington, DC: International City Management Association} {\bf
  1991}.

\bibitem[Cronstedt \em{et~al.}(2002)Cronstedt et~al.]{cronstedt2002prevention}
Cronstedt, M.; others.
\newblock Prevention, preparedness, response, recovery-an outdated concept?
\newblock {\em Australian Journal of Emergency Management, The} {\bf 2002},
  {\em 17},~10.

\bibitem[Haddow \em{et~al.}(2017)Haddow, Bullock, and
  Coppola]{haddow2017introduction}
Haddow, G.; Bullock, J.; Coppola, D.P.
\newblock {\em Introduction to emergency management}; Butterworth-Heinemann,
  2017.

\bibitem[Gelenbe and Cao(1998)]{Autonomous98}
Gelenbe, E.; Cao, Y.
\newblock Autonomous search for mines.
\newblock {\em European Journal of Operational Research} {\bf 1998}, {\em
  108},~319--333.

\bibitem[Gelenbe \em{et~al.}(2005)Gelenbe, Hussain, and Kaptan]{Kaptan}
Gelenbe, E.; Hussain, K.; Kaptan, V.
\newblock Autonomous search for mines.
\newblock {\em Journal of Systems and Software} {\bf 2005}, {\em 74},~255--268.

\bibitem[Belardo \em{et~al.}(1984)Belardo, Karwan, and
  Wallace]{belardo1984investigation}
Belardo, S.; Karwan, K.R.; Wallace, W.
\newblock An investigation of system design considerations for emergency
  management decision support.
\newblock {\em Systems, Man and Cybernetics, IEEE Transactions on} {\bf 1984},
  pp. 795--804.

\bibitem[Chalmet \em{et~al.}(1982)Chalmet, Francis, and
  Saunders]{chalmet1982network}
Chalmet, L.; Francis, R.; Saunders, P.
\newblock Network models for building evacuation.
\newblock {\em Fire Technology} {\bf 1982}, {\em 18},~90--113.

\bibitem[Hughes(1990)]{hughes1990graph}
Hughes, T.
\newblock Graph processing by which to evacuate a mine.
\newblock  Applied Computing, 1990., Proceedings of the 1990 Symposium on,
  1990, pp. 137--141.

\bibitem[Southworth \em{et~al.}(1989)Southworth, Chin, and
  Cheng]{southworth1989road}
Southworth, F.; Chin, S.M.; Cheng, P.
\newblock A telemetric monitoring and analysis system for use during large
  scale population evacuations.
\newblock  Road Traffic Monitoring, 1989., Second International Conference on,
  1989, pp. 99--103.

\bibitem[Griffith(1982)]{griffith1982hurricane}
Griffith, D.
\newblock The critical problems of hurricane evacuation and alternative
  solutions.
\newblock  OCEANS 82,  1982, pp. 990--994.

\bibitem[Zorpette(1987)]{zorpette1987planning}
Zorpette, G.
\newblock Evacuation planning for Lilco's Shoreham plant: Lack of an approved
  emergency evacuation strategy may prevent full operation of a \$5 billion
  nuclear power plant on long Island's north shore.
\newblock {\em Spectrum, IEEE} {\bf 1987}, {\em 24},~22--24.

\bibitem[Serpa \em{et~al.}(1981)Serpa, Walker, and Jenckes]{serpa1981emergency}
Serpa, D.P.; Walker, D.M.; Jenckes, T.A.
\newblock Emergency Monitoring, Assessment and Response System for Diablo
  Canyon Nuclear Power Plant.
\newblock {\em Nuclear Science, IEEE Transactions on} {\bf 1981}, {\em
  28},~236--241.

\bibitem[Filippoupolitis and Gelenbe(2009)]{filippoupolitis2009distributed}
Filippoupolitis, A.; Gelenbe, E.
\newblock A distributed decision support system for Building Evacuation.
\newblock  Human System Interactions, 2009. HSI '09. 2nd Conference on,  2009,
  pp. 323--330.

\bibitem[Tseng \em{et~al.}(2006)Tseng, Pan, and Tsai]{tseng2006wireless}
Tseng, Y.C.; Pan, M.S.; Tsai, Y.Y.
\newblock Wireless sensor networks for emergency navigation.
\newblock {\em Computer} {\bf 2006}, {\em 39},~55--62.

\bibitem[Park and Corson(1997)]{park1997highly}
Park, V.D.; Corson, M.S.
\newblock A highly adaptive distributed routing algorithm for mobile wireless
  networks.
\newblock  INFOCOM'97. Sixteenth Annual Joint Conference of the IEEE Computer
  and Communications Societies. Driving the Information Revolution.,
  Proceedings IEEE. IEEE,  1997, Vol.~3, pp. 1405--1413.

\bibitem[Li \em{et~al.}(2003)Li, De~Rosa, and Rus]{li2003distributed}
Li, Q.; De~Rosa, M.; Rus, D.
\newblock Distributed algorithms for guiding navigation across a sensor
  network.
\newblock  Proceedings of the 9th annual international conference on Mobile
  computing and networking. ACM,  2003, pp. 313--325.

\bibitem[Koditschek(1989)]{koditschek1989robot}
Koditschek, D.E.
\newblock Robot planning and control via potential functions.
\newblock {\em The robotics review} {\bf 1989}, p. 349.

\bibitem[Gorbil and Gelenbe(2011)]{gorbil2011opportunistic}
Gorbil, G.; Gelenbe, E.
\newblock Opportunistic communications for emergency support systems.
\newblock {\em Procedia Computer Science} {\bf 2011}, {\em 5},~39--47.

\bibitem[Gelenbe and Gorbil(2012)]{gelenbe2012wireless}
Gelenbe, E.; Gorbil, G.
\newblock Wireless networks in emergency management.
\newblock  Proceedings of the first ACM international workshop on Practical
  issues and applications in next generation wireless networks. ACM,  2012, pp.
  1--6.

\bibitem[Gorbil and Gelenbe(2013)]{gorbil2013resilient}
Gorbil, G.; Gelenbe, E.
\newblock Resilient emergency evacuation using opportunistic communications. In
  {\em Computer and Information Sciences III}; Springer,  2013; pp. 249--257.

\bibitem[Pelusi \em{et~al.}(2006)Pelusi, Passarella, and
  Conti]{pelusi2006opportunistic}
Pelusi, L.; Passarella, A.; Conti, M.
\newblock Opportunistic networking: data forwarding in disconnected mobile ad
  hoc networks.
\newblock {\em Communications Magazine, IEEE} {\bf 2006}, {\em 44},~134--141.

\bibitem[Filippoupolitis \em{et~al.}(2011)Filippoupolitis, Gorbil, and
  Gelenbe]{filippoupolitis2011autonomous}
Filippoupolitis, A.; Gorbil, G.; Gelenbe, E.
\newblock Autonomous navigation systems for emergency management in buildings.
\newblock  GLOBECOM Workshops (GC Wkshps), 2011 IEEE,  2011, pp. 1056--1061.

\bibitem[Zubair \em{et~al.}(2011)Zubair, bnt Fisal, Yerima, Salihu, and
  Salihu]{zubair2011adaptive}
Zubair, S.; bnt Fisal, N.; Yerima, S.; Salihu, B.; Salihu, Y.
\newblock CoWiSMoN: A framework for cognitive wireless sensor mobile network
  system for emergency rescue management.
\newblock  Adaptive Science and Technology (ICAST), 2011 3rd IEEE International
  Conference,  2011, pp. 237--241.

\bibitem[Inoue \em{et~al.}(2008)Inoue, Sashima, Ikeda, and
  Kurumatani]{inoue2008indoor}
Inoue, Y.; Sashima, A.; Ikeda, T.; Kurumatani, K.
\newblock Indoor Emergency Evacuation Service on Autonomous Navigation System
  using Mobile Phone.
\newblock  Universal Communication, 2008. ISUC '08. Second International
  Symposium on,  2008, pp. 79--85.

\bibitem[Tseng \em{et~al.}(2007)Tseng, Wang, Cheng, and Hsieh]{tseng2007imouse}
Tseng, Y.C.; Wang, Y.C.; Cheng, K.Y.; Hsieh, Y.Y.
\newblock iMouse: An Integrated Mobile Surveillance and Wireless Sensor System.
\newblock {\em Computer} {\bf 2007}, {\em 40},~60--66.

\bibitem[Dong \em{et~al.}(2014)Dong, Li, Ota, Yang, and
  Zhu]{dong2014multicloud}
Dong, M.; Li, H.; Ota, K.; Yang, L.; Zhu, H.
\newblock Multicloud-Based Evacuation Services for Emergency Management.
\newblock {\em Cloud Computing, IEEE} {\bf 2014}, {\em 1},~50--59.

\bibitem[Chu and Wu(2011)]{chu2011integrated}
Chu, L.; Wu, S.J.
\newblock An Integrated Building Fire Evacuation System with RFID and Cloud
  Computing.
\newblock  Intelligent Information Hiding and Multimedia Signal Processing
  (IIH-MSP), 2011 Seventh International Conference on,  2011, pp. 17--20.

\bibitem[Qiu \em{et~al.}(2014)Qiu, Ming, Wang, Yang, and Xiang]{qiu2014cloud}
Qiu, M.; Ming, Z.; Wang, J.; Yang, L.T.; Xiang, Y.
\newblock Enabling Cloud Computing in Emergency Management Systems.
\newblock {\em IEEE Cloud Computing} {\bf 2014}, {\em 1},~60--67.

\bibitem[Gelenbe and Bi(2014)]{gelenbebi2014Emergency}
Gelenbe, E.; Bi, H.
\newblock Emergency Navigation without an Infrastructure.
\newblock {\em Sensors} {\bf 2014}, {\em 14},~15142--15162.

\bibitem[Helbing \em{et~al.}(2000)Helbing, Farkas, and Vicsek]{Helbing1}
Helbing, D.; Farkas, I.; Vicsek, T.
\newblock Simulating dynamical features of escape panic.
\newblock {\em Nature} {\bf 2000}, {\em 407},~487--490.

\bibitem[Yang \em{et~al.}(2005)Yang, Zhao, Li, and Fang]{yang2005simulation}
Yang, L.; Zhao, D.; Li, J.; Fang, T.
\newblock Simulation of the kin behavior in building occupant evacuation based
  on cellular automaton.
\newblock {\em Building and Environment} {\bf 2005}, {\em 40},~411--415.

\bibitem[Gwynne \em{et~al.}(1999)Gwynne, Galea, Owen, Lawrence, Filippidis,
  et~al.]{gwynne1999review}
Gwynne, S.; Galea, E.; Owen, M.; Lawrence, P.J.; Filippidis, L.; others.
\newblock A review of the methodologies used in evacuation modelling.
\newblock {\em Fire and Materials} {\bf 1999}, {\em 23},~383--388.

\bibitem[Zheng \em{et~al.}(2009)Zheng, Zhong, and Liu]{zheng2009modeling}
Zheng, X.; Zhong, T.; Liu, M.
\newblock Modeling crowd evacuation of a building based on seven methodological
  approaches.
\newblock {\em Building and Environment} {\bf 2009}, {\em 44},~437--445.

\bibitem[Wolfram(1983)]{wolfram1983statistical}
Wolfram, S.
\newblock Statistical mechanics of cellular automata.
\newblock {\em Reviews of modern physics} {\bf 1983}, {\em 55},~601.

\bibitem[Bandini \em{et~al.}(2004)Bandini, Manzoni, and Vizzari]{Bandini1}
Bandini, S.; Manzoni, S.; Vizzari, G.
\newblock Situated Cellular Agents: a model to simulate crowding dynamics.
\newblock {\em IEICE Transactions on Information and Systems: Special Issue on
  Cellular Automata} {\bf 2004}, {\em E87-D(3)},~669--676.

\bibitem[Yu and Song(2007)]{yu2007cellular}
Yu, Y.; Song, W.
\newblock Cellular automaton simulation of pedestrian counter flow considering
  the surrounding environment.
\newblock {\em Physical Review E} {\bf 2007}, {\em 75},~046112.

\bibitem[Spartalis \em{et~al.}(2014)Spartalis, Georgoudas, and
  Sirakoulis]{spartalis2014crowd}
Spartalis, E.; Georgoudas, I.G.; Sirakoulis, G.C.
\newblock CA Crowd Modeling for a Retirement House Evacuation with Guidance. In
  {\em Cellular Automata}; Springer,  2014; pp. 481--491.

\bibitem[M{\"u}ller \em{et~al.}(2014)M{\"u}ller, Wohak, and
  Schadschneider]{muller2014study}
M{\"u}ller, F.; Wohak, O.; Schadschneider, A.
\newblock Study of Influence of Groups on Evacuation Dynamics Using a Cellular
  Automaton Model.
\newblock {\em Transportation Research Procedia} {\bf 2014}, {\em 2},~168--176.

\bibitem[Helbing and Molnar(1995)]{helbing1995social}
Helbing, D.; Molnar, P.
\newblock Social force model for pedestrian dynamics.
\newblock {\em Physical review E} {\bf 1995}, {\em 51},~4282.

\bibitem[Parisi and Dorso(2005)]{parisi2005microscopic}
Parisi, D.; Dorso, C.
\newblock Microscopic dynamics of pedestrian evacuation.
\newblock {\em Physica A: Statistical Mechanics and its Applications} {\bf
  2005}, {\em 354},~606--618.

\bibitem[Seyfried \em{et~al.}(2006)Seyfried, Steffen, and
  Lippert]{seyfried2006basics}
Seyfried, A.; Steffen, B.; Lippert, T.
\newblock Basics of modelling the pedestrian flow.
\newblock {\em Physica A: Statistical Mechanics and its Applications} {\bf
  2006}, {\em 368},~232--238.

\bibitem[Henderson(1971)]{henderson1971statistics}
Henderson, L.
\newblock The statistics of crowd fluids.
\newblock {\em Nature} {\bf 1971}, {\em 229},~381--383.

\bibitem[Helbing \em{et~al.}(2002)Helbing, Farkas, Molnar, and
  Vicsek]{helbing2002simulation}
Helbing, D.; Farkas, I.J.; Molnar, P.; Vicsek, T.
\newblock Simulation of pedestrian crowds in normal and evacuation situations.
\newblock {\em Pedestrian and evacuation dynamics} {\bf 2002}, {\em
  21},~21--58.

\bibitem[Fredkin and Toffoli(2002)]{fredkin2002conservative}
Fredkin, E.; Toffoli, T.
\newblock {\em Conservative logic}; Springer,  2002.

\bibitem[Tajima \em{et~al.}(2001)Tajima, Takimoto, and
  Nagatani]{tajima2001scaling}
Tajima, Y.; Takimoto, K.; Nagatani, T.
\newblock Scaling of pedestrian channel flow with a bottleneck.
\newblock {\em Physica A: Statistical Mechanics and its Applications} {\bf
  2001}, {\em 294},~257--268.

\bibitem[Takimoto and Nagatani(2003)]{takimoto2003spatio}
Takimoto, K.; Nagatani, T.
\newblock Spatio-temporal distribution of escape time in evacuation process.
\newblock {\em Physica A: Statistical Mechanics and its Applications} {\bf
  2003}, {\em 320},~611--621.

\bibitem[Hoogendoorn and HL~Bovy(2003)]{hoogendoorn2003simulation}
Hoogendoorn, S.; HL~Bovy, P.
\newblock Simulation of pedestrian flows by optimal control and differential
  games.
\newblock {\em Optimal Control Applications and Methods} {\bf 2003}, {\em
  24},~153--172.

\bibitem[Lo \em{et~al.}(2006)Lo, Huang, Wang, and Yuen]{lo2006game}
Lo, S.M.; Huang, H.C.; Wang, P.; Yuen, K.
\newblock A game theory based exit selection model for evacuation.
\newblock {\em Fire Safety Journal} {\bf 2006}, {\em 41},~364--369.

\bibitem[Ehtamo \em{et~al.}(2010)Ehtamo, Heli{\"o}vaara, Hostikka, and
  Korhonen]{ehtamo2010modeling}
Ehtamo, H.; Heli{\"o}vaara, S.; Hostikka, S.; Korhonen, T.
\newblock Modeling evacuees’ exit selection with best response dynamics. In
  {\em Pedestrian and Evacuation Dynamics 2008}; Springer,  2010; pp. 309--319.

\bibitem[Zheng and Cheng(2011)]{zheng2011modeling}
Zheng, X.; Cheng, Y.
\newblock Modeling cooperative and competitive behaviors in emergency
  evacuation: A game-theoretical approach.
\newblock {\em Computers \& Mathematics with Applications} {\bf 2011}, {\em
  62},~4627--4634.

\bibitem[Bonabeau(2002)]{bonabeau2002agent}
Bonabeau, E.
\newblock Agent-based modeling: Methods and techniques for simulating human
  systems.
\newblock {\em Proceedings of the National Academy of Sciences of the United
  States of America} {\bf 2002}, {\em 99},~7280--7287.

\bibitem[Zarboutis and Marmaras(2004)]{zarboutis2004searching}
Zarboutis, N.; Marmaras, N.
\newblock Searching efficient plans for emergency rescue through simulation:
  the case of a metro fire.
\newblock {\em Cognition, Technology \& Work} {\bf 2004}, {\em 6},~117--126.

\bibitem[Goldstone and Janssen(2005)]{goldstone2005computational}
Goldstone, R.L.; Janssen, M.A.
\newblock Computational models of collective behavior.
\newblock {\em Trends in cognitive sciences} {\bf 2005}, {\em 9},~424--430.

\bibitem[Pan \em{et~al.}(2007)Pan, Han, Dauber, and Law]{pan2007multi}
Pan, X.; Han, C.S.; Dauber, K.; Law, K.H.
\newblock A multi-agent based framework for the simulation of human and social
  behaviors during emergency evacuations.
\newblock {\em Ai \& Society} {\bf 2007}, {\em 22},~113--132.

\bibitem[Saloma \em{et~al.}(2003)Saloma, Perez, Tapang, Lim, and
  Palmes-Saloma]{saloma2003self}
Saloma, C.; Perez, G.J.; Tapang, G.; Lim, M.; Palmes-Saloma, C.
\newblock Self-organized queuing and scale-free behavior in real escape panic.
\newblock {\em Proceedings of the National Academy of Sciences} {\bf 2003},
  {\em 100},~11947--11952.

\bibitem[Altshuler \em{et~al.}(2005)Altshuler, Ramos, N{\'u}{\~n}ez,
  Fern{\'a}ndez, Batista-Leyva, and Noda]{altshuler2005symmetry}
Altshuler, E.; Ramos, O.; N{\'u}{\~n}ez, Y.; Fern{\'a}ndez, J.; Batista-Leyva,
  A.; Noda, C.
\newblock Symmetry breaking in escaping ants.
\newblock {\em The American Naturalist} {\bf 2005}, {\em 166},~643--649.

\bibitem[Pelechano and Malkawi(2008)]{pelechano2008evacuation}
Pelechano, N.; Malkawi, A.
\newblock Evacuation simulation models: Challenges in modeling high rise
  building evacuation with cellular automata approaches.
\newblock {\em Automation in construction} {\bf 2008}, {\em 17},~377--385.

\bibitem[Perez \em{et~al.}(2002)Perez, Tapang, Lim, and
  Saloma]{perez2002streaming}
Perez, G.J.; Tapang, G.; Lim, M.; Saloma, C.
\newblock Streaming, disruptive interference and power-law behavior in the exit
  dynamics of confined pedestrians.
\newblock {\em Physica A: Statistical Mechanics and its Applications} {\bf
  2002}, {\em 312},~609--618.

\bibitem[Burstedde \em{et~al.}(2004)Burstedde, Klauck, Schadschneider, and
  Zittartz]{Burstedde1}
Burstedde, C.; Klauck, K.; Schadschneider, A.; Zittartz, J.
\newblock Simulation of pedestrian dynamics using a two-dimensional cellular
  automaton.
\newblock {\em Physica A} {\bf 2004}, {\em 295},~507--525.

\bibitem[Zheng \em{et~al.}(2002)Zheng, Kashimori, and
  Kambara]{zheng2002collective}
Zheng, M.; Kashimori, Y.; Kambara, T.
\newblock A model describing collective behaviors of pedestrians with various
  personalities in danger situations.
\newblock  Neural Information Processing, 2002. ICONIP '02. Proceedings of the
  9th International Conference on,  2002, Vol.~4, pp. 2083--2087 vol.4.

\bibitem[Hecht-Nielsen(1988)]{hecht1988applications}
Hecht-Nielsen, R.
\newblock Applications of counterpropagation networks.
\newblock {\em Neural networks} {\bf 1988}, {\em 1},~131--139.

\bibitem[Hughes(2002)]{hughes2002continuum}
Hughes, R.L.
\newblock A continuum theory for the flow of pedestrians.
\newblock {\em Transportation Research Part B: Methodological} {\bf 2002}, {\em
  36},~507--535.

\bibitem[Colombo and Rosini(2005)]{colombo2005pedestrian}
Colombo, R.M.; Rosini, M.D.
\newblock Pedestrian flows and non-classical shocks.
\newblock {\em Mathematical Methods in the Applied Sciences} {\bf 2005}, {\em
  28},~1553--1567.

\bibitem[Lighthill and Whitham(1955)]{lighthill1955kinematic}
Lighthill, M.J.; Whitham, G.B.
\newblock On kinematic waves. II. A theory of traffic flow on long crowded
  roads.
\newblock  Proceedings of the Royal Society of London A: Mathematical, Physical
  and Engineering Sciences. The Royal Society,  1955, Vol. 229, pp. 317--345.

\bibitem[Liu(1975)]{liu1975riemann}
Liu, T.P.
\newblock The Riemann problem for general systems of conservation laws.
\newblock {\em Journal of Differential Equations} {\bf 1975}, {\em
  18},~218--234.

\bibitem[Thouless(2013)]{thouless2013quantum}
Thouless, D.J.
\newblock {\em The quantum mechanics of many-body systems}; Courier
  Corporation,  2013.

\bibitem[Kirchner \em{et~al.}(2003)Kirchner, Kl{\"u}pfel, Nishinari,
  Schadschneider, and Schreckenberg]{kirchner2003simulation}
Kirchner, A.; Kl{\"u}pfel, H.; Nishinari, K.; Schadschneider, A.;
  Schreckenberg, M.
\newblock Simulation of competitive egress behavior: comparison with aircraft
  evacuation data.
\newblock {\em Physica A: Statistical Mechanics and its Applications} {\bf
  2003}, {\em 324},~689--697.

\bibitem[Fudenberg and Tirole(1991)]{fudenberg1991game}
Fudenberg, D.; Tirole, J.
\newblock Game theory, 1991.
\newblock {\em Cambridge, Massachusetts} {\bf 1991}, {\em 393}.

\bibitem[Nash(1951)]{nash1951non}
Nash, J.
\newblock Non-cooperative games.
\newblock {\em Annals of mathematics} {\bf 1951}, pp. 286--295.

\bibitem[Nakamura and Asada(1995)]{nakamura1995motion}
Nakamura, T.; Asada, M.
\newblock Motion sketch: Acquisition of visual motion guided behaviors.
\newblock  IJCAI,  1995, Vol.~95, pp. 126--132.

\bibitem[March(1994)]{march1994primer}
March, J.G.
\newblock {\em Primer on decision making: How decisions happen}; Simon and
  Schuster,  1994.

\bibitem[Kuligowski \em{et~al.}(2005)Kuligowski, Peacock, and
  Hoskins]{kuligowski2005review}
Kuligowski, E.D.; Peacock, R.D.; Hoskins, B.
\newblock {\em A review of building evacuation models}; US Department of
  Commerce, National Institute of Standards and Technology,  2005.

\bibitem[Gelenbe and Wu(2012)]{GelenbeW12}
Gelenbe, E.; Wu, F.J.
\newblock Large scale simulation for human evacuation and rescue.
\newblock {\em Computers {\&} Mathematics with Applications} {\bf 2012}, {\em
  64},~3869--3880.

\bibitem[Gelenbe and Wu(2013)]{GelenbeW13}
Gelenbe, E.; Wu, F.J.
\newblock Future research on cyber-physical emergency management systems.
\newblock {\em Future Internet} {\bf 2013}, {\em 5},~336--354.

\bibitem[Francis(1984)]{francis1984negative}
Francis, R.
\newblock {\em A negative exponential solution to an evacuation problem};
  National Bureau of Standards, Center for Fire Research,  1984.

\bibitem[Kisko and Francis(1985)]{kisko1985evacnet}
Kisko, T.M.; Francis, R.L.
\newblock EVACNET+: a computer program to determine optimal building evacuation
  plans.
\newblock {\em Fire Safety Journal} {\bf 1985}, {\em 9},~211--220.

\bibitem[Lu \em{et~al.}(2003)Lu, Huang, and Shekhar]{lu2003evacuation}
Lu, Q.; Huang, Y.; Shekhar, S.
\newblock Evacuation planning: a capacity constrained routing approach. In {\em
  Intelligence and Security Informatics}; Springer,  2003; pp. 111--125.

\bibitem[Lu \em{et~al.}(2005)Lu, George, and Shekhar]{lu2005capacity}
Lu, Q.; George, B.; Shekhar, S.
\newblock Capacity constrained routing algorithms for evacuation planning: A
  summary of results. In {\em Advances in spatial and temporal databases};
  Springer,  2005; pp. 291--307.

\bibitem[Chen \em{et~al.}(2008)Chen, Chen, and Shen]{chen2008distributed}
Chen, P.Y.; Chen, W.T.; Shen, Y.T.
\newblock A distributed area-based guiding navigation protocol for wireless
  sensor networks.
\newblock  Parallel and Distributed Systems, 2008. ICPADS'08. 14th IEEE
  International Conference on. IEEE,  2008, pp. 647--654.

\bibitem[Wang \em{et~al.}(2013)Wang, Li, Li, Liu, and Yang]{wang2013sensor}
Wang, J.; Li, Z.; Li, M.; Liu, Y.; Yang, Z.
\newblock Sensor network navigation without locations.
\newblock {\em Parallel and Distributed Systems, IEEE Transactions on} {\bf
  2013}, {\em 24},~1436--1446.

\bibitem[MacGregor~Smith(1991)]{macgregor1991state}
MacGregor~Smith, J.
\newblock State-dependent queueing models in emergency evacuation networks.
\newblock {\em Transportation Research Part B: Methodological} {\bf 1991}, {\em
  25},~373--389.

\bibitem[Cruz \em{et~al.}(2005)Cruz, Smith, and Queiroz]{cruz2005service}
Cruz, F.R.; Smith, J.M.; Queiroz, D.
\newblock Service and capacity allocation in M/G/c/c state-dependent queueing
  networks.
\newblock {\em Computers \& operations research} {\bf 2005}, {\em
  32},~1545--1563.

\bibitem[Stepanov and Smith(2009)]{stepanov2009multi}
Stepanov, A.; Smith, J.M.
\newblock Multi-objective evacuation routing in transportation networks.
\newblock {\em European Journal of Operational Research} {\bf 2009}, {\em
  198},~435--446.

\bibitem[Lino \em{et~al.}(2009)Lino, Maione, and Maione]{lino2009modeling}
Lino, P.; Maione, G.; Maione, B.
\newblock Modeling and simulation of crowd egress dynamics in a discrete event
  environment.
\newblock  Control Applications,(CCA) \& Intelligent Control,(ISIC), 2009 IEEE.
  IEEE,  2009, pp. 843--848.

\bibitem[Desmet and Gelenbe(2013)]{desmet2013graph}
Desmet, A.; Gelenbe, E.
\newblock Graph and analytical models for emergency evacuation.
\newblock {\em Future Internet} {\bf 2013}, {\em 5},~46--55.

\bibitem[Bi(2016)]{huibo2016flowoptimisation}
Bi, H.
\newblock Evacuee Flow Optimisation Using G-Network with Multiple Classes of
  Positive Customers.
\newblock  2016 IEEE 24th International Symposium on Modeling, Analysis and
  Simulation of Computer and Telecommunication Systems (MASCOTS),  2016, pp.
  135--143.

\bibitem[Bi and Abdelrahman(2016)]{bi2016energyaware}
Bi, H.; Abdelrahman, O.H.
\newblock ENERGY-AWARE NAVIGATION IN LARGE-SCALE EVACUATION USING G-NETWORKS.
\newblock {\em Probability in the Engineering and Informational Sciences} {\bf
  2016}, pp. 1--13.

\bibitem[Chen \em{et~al.}(2008)Chen, Chen, Wu, and Huang]{chen2008load}
Chen, W.T.; Chen, P.Y.; Wu, C.H.; Huang, C.F.
\newblock A load-balanced guiding navigation protocol in wireless sensor
  networks.
\newblock  Global Telecommunications Conference, 2008. IEEE GLOBECOM 2008.
  IEEE. IEEE,  2008, pp. 1--6.

\bibitem[Gelenbe and Timotheou(2008)]{RNN2008}
Gelenbe, E.; Timotheou, S.
\newblock Random neural networks with synchronized interactions.
\newblock {\em Neural Computation} {\bf 2008}, {\em 20},~2308--2324.

\bibitem[Jankowska \em{et~al.}(2009)Jankowska, Schut, and
  Ferreira-Schut]{jankowska2009wireless}
Jankowska, A.; Schut, M.; Ferreira-Schut, N.
\newblock A wireless actuator-sensor neural network for evacuation routing.
\newblock  Sensor Technologies and Applications, 2009. SENSORCOMM'09. Third
  International Conference on. IEEE,  2009, pp. 139--144.

\bibitem[Li \em{et~al.}(2010)Li, Fang, Li, and Zong]{li2010multiobjective}
Li, Q.; Fang, Z.; Li, Q.; Zong, X.
\newblock Multiobjective evacuation route assignment model based on genetic
  algorithm.
\newblock  Geoinformatics, 2010 18th International Conference on. IEEE,  2010,
  pp. 1--5.

\bibitem[Filippoupolitis(2010)]{filippoupolitis2010emergency}
Filippoupolitis, A.
\newblock Emergency Simulation and Decision Support Algorithms.
\newblock PhD thesis, Imperial College London (University of London),  2010.

\bibitem[Bi \em{et~al.}(2013)Bi, Desmet, and Gelenbe]{BiDesmetGelenbeISCIS2013}
Bi, H.; Desmet, A.; Gelenbe, E.
\newblock Routing Emergency Evacuees with Cognitive Packet Networks.
\newblock  Proceedings of the 28th International Symposium on Computer and
  Information Sciences ({ISCIS'13}). Springer, London,  2013.

\bibitem[Bi and Gelenbe(2014)]{biandgelenberouting2014}
Bi, H.; Gelenbe, E.
\newblock Routing diverse evacuees with Cognitive Packets.
\newblock  Pervasive Computing and Communications Workshops (PERCOM Workshops),
  2014 IEEE International Conference on,  2014, pp. 291--296.

\bibitem[Hasofer and Odigie(2001)]{hasofer2001stochastic}
Hasofer, A.; Odigie, D.
\newblock Stochastic modelling for occupant safety in a building fire.
\newblock {\em Fire Safety Journal} {\bf 2001}, {\em 36},~269--289.

\bibitem[Barnes \em{et~al.}(2007)Barnes, Leather, and
  Arvind]{barnes2007emergency}
Barnes, M.; Leather, H.; Arvind, D.
\newblock Emergency evacuation using wireless sensor networks.
\newblock  Local Computer Networks, 2007. LCN 2007. 32nd IEEE Conference on.
  IEEE,  2007, pp. 851--857.

\bibitem[Han \em{et~al.}(2010)Han, Potter, Beckett, Pringle, Welch, Koo,
  Wickler, Usmani, Torero, and Tate]{han2010firegrid}
Han, L.; Potter, S.; Beckett, G.; Pringle, G.; Welch, S.; Koo, S.H.; Wickler,
  G.; Usmani, A.; Torero, J.L.; Tate, A.
\newblock FireGrid: an e-infrastructure for next-generation emergency response
  support.
\newblock {\em Journal of Parallel and Distributed Computing} {\bf 2010}, {\em
  70},~1128--1141.

\bibitem[Radianti \em{et~al.}(2015)Radianti, Granmo, Sarshar, Goodwin, Dugdale,
  and Gonzalez]{radianti2015spatio}
Radianti, J.; Granmo, O.C.; Sarshar, P.; Goodwin, M.; Dugdale, J.; Gonzalez,
  J.J.
\newblock A spatio-temporal probabilistic model of hazard-and crowd dynamics
  for evacuation planning in disasters.
\newblock {\em Applied Intelligence} {\bf 2015}, {\em 42},~3--23.

\bibitem[Bi and Gelenbe(2015)]{BIGELENBEIEEEPERCOM2015}
Bi, H.; Gelenbe, E.
\newblock Cloud enabled emergency navigation using faster-than-real-time
  simulation.
\newblock  2015 IEEE International Conference on Pervasive Computing and
  Communication Workshops (PerCom Workshops),  2015, pp. 475--480.

\bibitem[Ford and Fulkerson(2010)]{ford2010flows}
Ford, D.; Fulkerson, D.R.
\newblock {\em Flows in networks}; Princeton university press,  2010.

\bibitem[Ahuja \em{et~al.}(1988)Ahuja, Magnanti, and Orlin]{ahuja1988network}
Ahuja, R.K.; Magnanti, T.L.; Orlin, J.B.
\newblock Network flows {\bf 1988}.

\bibitem[Graves \em{et~al.}(1977)Graves, Brown, and Bradley]{graves1977design}
Graves, G.W.; Brown, G.G.; Bradley, G.H.
\newblock Design and Implementation of Large-Scale Primal Transshipment
  Algorithms.
\newblock {\em Management Science} {\bf 1977}, {\em 24},~1--34.

\bibitem[Hoppe and Tardos(1994)]{hoppe1994polynomial}
Hoppe, B.; Tardos, {\'E}.
\newblock Polynomial time algorithms for some evacuation problems.
\newblock  Proceedings of the fifth annual ACM-SIAM symposium on Discrete
  algorithms. Society for Industrial and Applied Mathematics,  1994, pp.
  433--441.

\bibitem[Hoppe and Tardos(2000)]{hoppe2000quickest}
Hoppe, B.; Tardos, {\'E}.
\newblock The quickest transshipment problem.
\newblock {\em Mathematics of Operations Research} {\bf 2000}, {\em
  25},~36--62.

\bibitem[Megiddo(1974)]{megiddo1974optimal}
Megiddo, N.
\newblock Optimal flows in networks with multiple sources and sinks.
\newblock {\em Mathematical Programming} {\bf 1974}, {\em 7},~97--107.

\bibitem[Minieka(1973)]{minieka1973maximal}
Minieka, E.
\newblock Maximal, lexicographic, and dynamic network flows.
\newblock {\em Operations Research} {\bf 1973}, {\em 21},~517--527.

\bibitem[Dijkstra(1959)]{dijkstra1959note}
Dijkstra, E.W.
\newblock A note on two problems in connexion with graphs.
\newblock {\em Numerische mathematik} {\bf 1959}, {\em 1},~269--271.

\bibitem[Li \em{et~al.}(2002)Li, Calinescu, and Wan]{li2002distributed}
Li, X.Y.; Calinescu, G.; Wan, P.J.
\newblock Distributed construction of a planar spanner and routing for ad hoc
  wireless networks.
\newblock  INFOCOM 2002. Twenty-First Annual Joint Conference of the IEEE
  Computer and Communications Societies. Proceedings. IEEE. IEEE,  2002,
  Vol.~3, pp. 1268--1277.

\bibitem[Li \em{et~al.}(2003)Li, Calinescu, Wan, and Wang]{li2003localized}
Li, X.Y.; Calinescu, G.; Wan, P.J.; Wang, Y.
\newblock Localized delaunay triangulation with application in ad hoc wireless
  networks.
\newblock {\em Parallel and Distributed Systems, IEEE Transactions on} {\bf
  2003}, {\em 14},~1035--1047.

\bibitem[Bruck \em{et~al.}(2007)Bruck, Gao, and Jiang]{bruck2007map}
Bruck, J.; Gao, J.; Jiang, A.
\newblock MAP: Medial axis based geometric routing in sensor networks.
\newblock {\em Wireless Networks} {\bf 2007}, {\em 13},~835--853.

\bibitem[Gelenbe and Muntz(1976)]{Muntz}
Gelenbe, E.; Muntz, R.R.
\newblock Probabilistic models of computer systems, Part I (Exact Results).
\newblock {\em Acta Informatica} {\bf 1976}, {\em 7},~35--60.

\bibitem[Gelenbe(1973)]{Unified}
Gelenbe, E.
\newblock A unified approach to a class of page replacement algorithms.
\newblock {\em IEEE Transactions on Computers} {\bf 1973}, {\em 22},~611--618.

\bibitem[Fruin(1971)]{fruin1971pedestrian}
Fruin, J.J.
\newblock {\em Pedestrian planning and design}; Metropolitan Association of
  Urban Designers and Environmental Planners, Inc.,  1971.

\bibitem[Tregenza(1976)]{tregenza1976design}
Tregenza, P.
\newblock {\em The design of interior circulation}; Van Nostrand Reinhold,
  1976.

\bibitem[Francis and Chalmet(1980)]{francis1980network}
Francis, R.; Chalmet, L.
\newblock {\em Network models for building evacuation: A prototype primer};
  1980.

\bibitem[Yuhaski~Jr and Smith(1989)]{yuhaski1989modeling}
Yuhaski~Jr, S.J.; Smith, J.M.
\newblock Modeling circulation systems in buildings using state dependent
  queueing models.
\newblock {\em Queueing Systems} {\bf 1989}, {\em 4},~319--338.

\bibitem[Cheah and Smith(1994)]{cheah1994generalized}
Cheah, J.Y.; Smith, J.M.
\newblock Generalized M/G/c/c state dependent queueing models and pedestrian
  traffic flows.
\newblock {\em Queueing Systems} {\bf 1994}, {\em 15},~365--386.

\bibitem[Kerbachea and Smith(1987)]{kerbachea1987generalized}
Kerbachea, L.; Smith, J.M.
\newblock The generalized expansion method for open finite queueing networks.
\newblock {\em European Journal of Operational Research} {\bf 1987}, {\em
  32},~448--461.

\bibitem[Kerbache and Smith(1988)]{kerbache1988asymptotic}
Kerbache, L.; Smith, J.M.
\newblock Asymptotic behavior of the expansion method for open finite queueing
  networks.
\newblock {\em Computers \& Operations Research} {\bf 1988}, {\em
  15},~157--169.

\bibitem[Smith \em{et~al.}(2000)Smith, Gershwin, and
  Papadopoulos]{smith2000performance}
Smith, J.M.; Gershwin, S.B.; Papadopoulos, C.T.
\newblock {\em Performance evaluation and optimization of production lines};
  Vol.~93, Baltzer Science Publishers,  2000.

\bibitem[Lino \em{et~al.}(2011)Lino, Pizzileo, Maione, and
  Maione]{lino2011tuning}
Lino, P.; Pizzileo, B.; Maione, G.; Maione, B.
\newblock Tuning and Validation of a Discrete-Event Model of the Egress
  Dynamics from Buildings*.
\newblock  World Congress,  2011, Vol.~18, pp. 8743--8748.

\bibitem[Helbing \em{et~al.}(2000)Helbing, Farkas, and
  Vicsek]{helbing2000simulating}
Helbing, D.; Farkas, I.; Vicsek, T.
\newblock Simulating dynamical features of escape panic.
\newblock {\em Nature} {\bf 2000}, {\em 407},~487--490.

\bibitem[Wang \em{et~al.}(2008)Wang, Luh, Chang, and Sun]{wang2008modeling}
Wang, P.; Luh, P.B.; Chang, S.C.; Sun, J.
\newblock Modeling and optimization of crowd guidance for building emergency
  evacuation.
\newblock  Automation Science and Engineering, 2008. CASE 2008. IEEE
  International Conference on. IEEE,  2008, pp. 328--334.

\bibitem[Watts(1987)]{watts1987computer}
Watts, J.M.
\newblock Computer models for evacuation analysis.
\newblock {\em Fire Safety Journal} {\bf 1987}, {\em 12},~237--245.

\bibitem[Cruz \em{et~al.}(2005)Cruz, Smith, and Medeiros]{cruz2005m}
Cruz, F.R.; Smith, J.M.; Medeiros, R.
\newblock An M/G/C/C state-dependent network simulation model.
\newblock {\em Computers \& Operations Research} {\bf 2005}, {\em
  32},~919--941.

\bibitem[Gelenbe and Labed(1998)]{gelenbe1998multiple}
Gelenbe, E.; Labed, A.
\newblock G-networks with multiple classes of signals and positive customers.
\newblock {\em European journal of operational research} {\bf 1998}, {\em
  108},~293--305.

\bibitem[Gelenbe(1993)]{gelenbe1993g}
Gelenbe, E.
\newblock G-networks with triggered customer movement.
\newblock {\em Journal of Applied Probability} {\bf 1993}, pp. 742--748.

\bibitem[Hill \em{et~al.}(2000)Hill, Szewczyk, Woo, Hollar, Culler, and
  Pister]{hill2000system}
Hill, J.; Szewczyk, R.; Woo, A.; Hollar, S.; Culler, D.; Pister, K.
\newblock System architecture directions for networked sensors.
\newblock  ACM SIGOPS operating systems review. ACM,  2000, Vol.~34, pp.
  93--104.

\bibitem[Pan \em{et~al.}(2006)Pan, Tsai, and Tseng]{pan2006emergency}
Pan, M.S.; Tsai, C.H.; Tseng, Y.C.
\newblock Emergency guiding and monitoring applications in indoor 3D
  environments by wireless sensor networks.
\newblock {\em International Journal of Sensor Networks} {\bf 2006}, {\em
  1},~2--10.

\bibitem[Gelenbe \em{et~al.}(1997)Gelenbe, Schmajuk, Staddon, and
  Reif]{gelenbe1997autonomous}
Gelenbe, E.; Schmajuk, N.; Staddon, J.; Reif, J.
\newblock Autonomous search by robots and animals: A survey.
\newblock {\em Robotics and Autonomous Systems} {\bf 1997}, {\em 22},~23--34.

\bibitem[Mitchell(1997)]{mitchell1997machine}
Mitchell, T.M.
\newblock Machine learning. 1997.
\newblock {\em Burr Ridge, IL: McGraw Hill} {\bf 1997}, {\em 45}.

\bibitem[John(1992)]{john1992holland}
John, H.
\newblock Holland, Adaptation in natural and artificial systems,  1992.

\bibitem[Deb \em{et~al.}(2000)Deb, Agrawal, Pratap, and Meyarivan]{deb2000fast}
Deb, K.; Agrawal, S.; Pratap, A.; Meyarivan, T.
\newblock A fast elitist non-dominated sorting genetic algorithm for
  multi-objective optimization: NSGA-II.
\newblock {\em Lecture notes in computer science} {\bf 2000}, {\em
  1917},~849--858.

\bibitem[Eppstein(1998)]{eppstein1998finding}
Eppstein, D.
\newblock Finding the k shortest paths.
\newblock {\em SIAM Journal on computing} {\bf 1998}, {\em 28},~652--673.

\bibitem[Saadatseresht \em{et~al.}(2009)Saadatseresht, Mansourian, and
  Taleai]{saadatseresht2009evacuation}
Saadatseresht, M.; Mansourian, A.; Taleai, M.
\newblock Evacuation planning using multiobjective evolutionary optimization
  approach.
\newblock {\em European Journal of Operational Research} {\bf 2009}, {\em
  198},~305--314.

\bibitem[Pan \em{et~al.}(2005)Pan, Han, and Law]{pan2005multi}
Pan, X.; Han, C.S.; Law, K.H.
\newblock A multi-agent based simulation framework for the study of human and
  social behavior in egress analysis.
\newblock  Proceedings of the ASCE International Conference on Computing in
  Civil Engineering,  2005, Vol.~92.

\bibitem[Samadzadegan and Yadegari()]{samadzadeganbiologically}
Samadzadegan, F.; Yadegari, M.
\newblock A BIOLOGICALLY-INSPIRED OPTIMIZATION ALGORITHM FOR URBAN EVACUATION
  PLANNING IN DISASTER MANAGEMENT.

\bibitem[Karaboga(2005)]{karaboga2005idea}
Karaboga, D.
\newblock An idea based on honey bee swarm for numerical optimization.
\newblock {\em Techn. Rep. TR06, Erciyes Univ. Press, Erciyes} {\bf 2005}.

\bibitem[Gelenbe \em{et~al.}(2001{\natexlab{a}})Gelenbe, Lent, and
  Xu]{gelenbe2001towards}
Gelenbe, E.; Lent, R.; Xu, Z.
\newblock Towards networks with cognitive packets. In {\em Performance and QoS
  of next generation networking}; Springer London,  2001; pp. 3--17.

\bibitem[Gelenbe \em{et~al.}(2001{\natexlab{b}})Gelenbe, Lent, and
  Xu]{gelenbe2001design}
Gelenbe, E.; Lent, R.; Xu, Z.
\newblock Design and performance of cognitive packet networks.
\newblock {\em Performance Evaluation} {\bf 2001}, {\em 46},~155--176.

\bibitem[Gelenbe(2003)]{gelenbe2003sensible}
Gelenbe, E.
\newblock Sensible decisions based on QoS.
\newblock {\em Computational management science} {\bf 2003}, {\em 1},~1--14.

\bibitem[Vahdat \em{et~al.}(2000)Vahdat, Becker, et~al.]{vahdat2000epidemic}
Vahdat, A.; Becker, D.; others.
\newblock Epidemic routing for partially connected ad hoc networks.
\newblock Technical report, Technical Report CS-200006, Duke University,  2000.

\bibitem[Gorbil and Gelenbe(2012)]{gorbil2012resilience}
Gorbil, G.; Gelenbe, E.
\newblock Resilience and security of opportunistic communications for emergency
  evacuation.
\newblock  Proceedings of the 7th ACM workshop on Performance monitoring and
  measurement of heterogeneous wireless and wired networks. ACM,  2012, pp.
  115--124.

\bibitem[Gelenbe(1993)]{gelenbe1993learning}
Gelenbe, E.
\newblock Learning in the recurrent random neural network.
\newblock {\em Neural Computation} {\bf 1993}, {\em 5},~154--164.

\bibitem[Desmet and Gelenbe(2014)]{Desmet2}
Desmet, A.; Gelenbe, E.
\newblock A Parametric Study of CPN's Convergence Process.
\newblock  Information Sciences and Systems 2014 - Proceedings of the 29th
  International Symposium on Computer and Information Sciences, ISCIS 2014,
  Krakow, Poland, October 27-28, 2014,  2014, pp. 13--20.

\bibitem[Bi(2014)]{birouting2014}
Bi, H.
\newblock Routing Diverse Evacuees with the Cognitive Packet Network Algorithm.
\newblock {\em Future Internet} {\bf 2014}, {\em 6},~203.

\bibitem[Akinwande \em{et~al.}(2015)Akinwande, Bi, and
  Gelenbe]{akinwandebigelenbedynamic2015}
Akinwande, O.; Bi, H.; Gelenbe, E.
\newblock Managing Crowds in Hazards With Dynamic Grouping.
\newblock {\em Access, IEEE} {\bf 2015}, {\em 3},~1060--1070.

\bibitem[Olenick and Carpenter(2003)]{olenick2003updated}
Olenick, S.M.; Carpenter, D.J.
\newblock An updated international survey of computer models for fire and
  smoke.
\newblock {\em Journal of Fire Protection Engineering} {\bf 2003}, {\em
  13},~87--110.

\bibitem[Koo \em{et~al.}(2008)Koo, Fraser-Mitchell, Upadhyay, and
  Welch]{koo2008sensor}
Koo, S.H.; Fraser-Mitchell, J.; Upadhyay, R.; Welch, S.
\newblock Sensor-linked fire simulation using a Monte-Carlo approach {\bf
  2008}.

\bibitem[Murphy(2002)]{murphy2002dynamic}
Murphy, K.P.
\newblock Dynamic bayesian networks: representation, inference and learning.
\newblock PhD thesis, University of California, Berkeley,  2002.

\bibitem[Gelenbe and Han(2014)]{gelenbe2014rescuerallocation}
Gelenbe, E.; Han, Q.
\newblock Near-Optimal Emergency Evacuation with Rescuer Allocation.
\newblock  In Proceedings of the 4th International Workshop on Pervasive
  Networks for Emergency Management (PerNEM'14). IEEE,  2014, pp. 1--6.

\bibitem[Rais \em{et~al.}(2009)Rais, Soh, Malek, Ahmad, Hashim, and
  Hall]{rais2009review}
Rais, N.; Soh, P.J.; Malek, F.; Ahmad, S.; Hashim, N.; Hall, P.
\newblock A review of wearable antenna.
\newblock  Antennas \& Propagation Conference, 2009. LAPC 2009. Loughborough.
  IEEE,  2009, pp. 225--228.

\bibitem[Orefice \em{et~al.}(2016)Orefice, Pirinoli, and
  Dassano]{orefice2016electrically}
Orefice, M.; Pirinoli, P.; Dassano, G.
\newblock Electrically-small wearable antennas for emergency services
  applications.
\newblock  Antenna Technology (iWAT), 2016 International Workshop on. IEEE,
  2016, pp. 131--134.

\bibitem[Lilja \em{et~al.}(2013)Lilja, Pyntt{\"a}ri, Kaija, M{\"a}kinen,
  Halonen, Sillanp{\"a}{\"a}, Heikkinen, M{\"a}ntysalo, Salonen, and
  de~Maagt]{lilja2013body}
Lilja, J.; Pyntt{\"a}ri, V.; Kaija, T.; M{\"a}kinen, R.; Halonen, E.;
  Sillanp{\"a}{\"a}, H.; Heikkinen, J.; M{\"a}ntysalo, M.; Salonen, P.;
  de~Maagt, P.
\newblock Body-worn antennas making a splash: Lifejacket-integrated antennas
  for global search and rescue satellite system.
\newblock {\em IEEE Antennas and Propagation Magazine} {\bf 2013}, {\em
  55},~324--341.

\bibitem[Li \em{et~al.}(2009)Li, Zhan, Wu, and Chen]{li2009ern}
Li, S.; Zhan, A.; Wu, X.; Chen, G.
\newblock ERN: Emergence rescue navigation with wireless sensor networks.
\newblock  Parallel and Distributed Systems (ICPADS), 2009 15th International
  Conference on. IEEE,  2009, pp. 361--368.

\bibitem[Huang \em{et~al.}(2005)Huang, Amjad, and Mishra]{huang2005CenWits}
Huang, J.H.; Amjad, S.; Mishra, S.
\newblock CenWits: A Sensor-based Loosely Coupled Search and Rescue System
  Using Witnesses.
\newblock  Proceedings of the 3rd International Conference on Embedded
  Networked Sensor Systems; ACM: New York, NY, USA,  2005; SenSys '05, pp.
  180--191.

\bibitem[Wu \em{et~al.}(2008)Wu, Cao, Zheng, and Zheng]{wu2008waiter}
Wu, W.; Cao, J.; Zheng, Y.; Zheng, Y.P.
\newblock WAITER: A Wearable Personal Healthcare and Emergency Aid System.
\newblock  2008 Sixth Annual IEEE International Conference on Pervasive
  Computing and Communications (PerCom),  2008, pp. 680--685.

\bibitem[Bozkurt \em{et~al.}(2014)Bozkurt, Roberts, Sherman, Brugarolas,
  Mealin, Majikes, Yang, and Loftin]{bozkurt2014toward}
Bozkurt, A.; Roberts, D.L.; Sherman, B.L.; Brugarolas, R.; Mealin, S.; Majikes,
  J.; Yang, P.; Loftin, R.
\newblock Toward cyber-enhanced working dogs for search and rescue.
\newblock {\em IEEE Intelligent Systems} {\bf 2014}, {\em 29},~32--39.

\bibitem[Bozkurt \em{et~al.}(2016)Bozkurt, Lobaton, and
  Sichitiu]{bozkurt2016biobotic}
Bozkurt, A.; Lobaton, E.; Sichitiu, M.
\newblock A biobotic distributed sensor network for under-rubble search and
  rescue.
\newblock {\em Computer} {\bf 2016}, {\em 49},~38--46.

\bibitem[Tadokoro(2009)]{satoshi2009earthquake}
Tadokoro, S.
\newblock Earthquake Disaster and Expectation for Robotics. In {\em Rescue
  Robotics}; Tadokoro, S., Ed.; Springer London,  2009; pp. 1--16.

\bibitem[Davids(2002)]{davids2002urban}
Davids, A.
\newblock Urban search and rescue robots: from tragedy to technology.
\newblock {\em IEEE Intelligent systems} {\bf 2002}, {\em 17},~81--83.

\bibitem[Barbera and Macintyre(1996)]{barbera1996urban}
Barbera, J.A.; Macintyre, A.
\newblock Urban search and rescue.
\newblock {\em Emergency medicine clinics of North America} {\bf 1996}, {\em
  14},~399--412.

\bibitem[Stopforth \em{et~al.}(2008)Stopforth, Holtzhausen, Bright, Tlale, and
  Kumile]{stopforth2008survey}
Stopforth, R.; Holtzhausen, S.; Bright, G.; Tlale, N.; Kumile, C.
\newblock Robots for Search and Rescue Purposes in Urban and Underwater
  Environments - a survey and comparison.
\newblock  Mechatronics and Machine Vision in Practice, 2008. M2VIP 2008. 15th
  International Conference on,  2008, pp. 476--480.

\bibitem[Casper and Murphy(2003)]{casper2003human}
Casper, J.; Murphy, R.R.
\newblock Human-robot interactions during the robot-assisted urban search and
  rescue response at the World Trade Center.
\newblock {\em Systems, Man, and Cybernetics, Part B: Cybernetics, IEEE
  Transactions on} {\bf 2003}, {\em 33},~367--385.

\bibitem[Doroodgar \em{et~al.}(2014)Doroodgar, Liu, and
  Nejat]{barzin2014learning}
Doroodgar, B.; Liu, Y.; Nejat, G.
\newblock A Learning-Based Semi-Autonomous Controller for Robotic Exploration
  of Unknown Disaster Scenes While Searching for Victims.
\newblock {\em IEEE Transactions on Cybernetics} {\bf 2014}, {\em
  44},~2719--2732.

\bibitem[Murphy(2011)]{murphy2011100}
Murphy, R.R.
\newblock The 100 : 100 challenge for computing in rescue robotics.
\newblock  Safety, Security, and Rescue Robotics (SSRR), 2011 IEEE
  International Symposium on. IEEE,  2011, pp. 72--75.

\bibitem[Hal{\'a}sz \em{et~al.}(2007)Hal{\'a}sz, Hsieh, Berman, and
  Kumar]{halasz2007dynamic}
Hal{\'a}sz, A.; Hsieh, M.A.; Berman, S.; Kumar, V.
\newblock Dynamic redistribution of a swarm of robots among multiple sites.
\newblock  Intelligent Robots and Systems, 2007. IROS 2007. IEEE/RSJ
  International Conference on. IEEE,  2007, pp. 2320--2325.

\bibitem[Berman \em{et~al.}(2009)Berman, Hal{\'a}sz, Hsieh, and
  Kumar]{berman2009optimized}
Berman, S.; Hal{\'a}sz, {\'A}.; Hsieh, M.A.; Kumar, V.
\newblock Optimized stochastic policies for task allocation in swarms of
  robots.
\newblock {\em IEEE Transactions on Robotics} {\bf 2009}, {\em 25},~927--937.

\bibitem[Iwano \em{et~al.}(2004)Iwano, Osuka, and Amano]{iwano2004rescuerobot}
Iwano, Y.; Osuka, K.; Amano, H.
\newblock Proposal of a Rescue Robot System in Nuclear-Power Plants -Rescue
  Activity via Small Vehicle Robots -.
\newblock  Robotics and Biomimetics, 2004. ROBIO 2004. IEEE International
  Conference on,  2004, pp. 227--232.

\bibitem[Ventura and Lima(2012)]{ventura2012search}
Ventura, R.; Lima, P.U.
\newblock Search and rescue robots: The civil protection teams of the future.
\newblock  Emerging Security Technologies (EST), 2012 Third International
  Conference on. IEEE,  2012, pp. 12--19.

\bibitem[Jacoff \em{et~al.}(2003)Jacoff, Messina, Weiss, Tadokoro, Nakagawa,
  et~al.]{jacoff2003test}
Jacoff, A.; Messina, E.; Weiss, B.; Tadokoro, S.; Nakagawa, Y.; others.
\newblock Test arenas and performance metrics for urban search and rescue
  robots.
\newblock  Intelligent Robots and Systems, 2003.(IROS 2003). Proceedings. 2003
  IEEE/RSJ International Conference on. IEEE,  2003, Vol.~4, pp. 3396--3403.

\bibitem[Ventura and Lima(2012)]{ventura2012future}
Ventura, R.; Lima, P.
\newblock Search and Rescue Robots: The Civil Protection Teams of the Future.
\newblock  Emerging Security Technologies (EST), 2012 Third International
  Conference on,  2012, pp. 12--19.

\bibitem[Tadokoro \em{et~al.}(2000)Tadokoro, Kitano, Takahashi, Noda,
  Matsubara, Joh, Koto, Takeuchi, Takahashi, Matsuno,
  et~al.]{tadokoro2000robocup}
Tadokoro, S.; Kitano, H.; Takahashi, T.; Noda, I.; Matsubara, H.; Joh, A.S.;
  Koto, T.; Takeuchi, I.; Takahashi, H.; Matsuno, F.; others.
\newblock The robocup-rescue project: A robotic approach to the disaster
  mitigation problem.
\newblock  Robotics and Automation, 2000. Proceedings. ICRA'00. IEEE
  International Conference on. IEEE,  2000, Vol.~4, pp. 4089--4094.

\bibitem[Gautam and Mohan(2012)]{gautam2012review}
Gautam, A.; Mohan, S.
\newblock A review of research in multi-robot systems.
\newblock  Industrial and Information Systems (ICIIS), 2012 7th IEEE
  International Conference on. IEEE,  2012, pp. 1--5.

\bibitem[Gelenbe and Cao(1998)]{gelenbe1998autonomous}
Gelenbe, E.; Cao, Y.
\newblock Autonomous search for mines.
\newblock {\em European Journal of Operational Research} {\bf 1998}, {\em
  108},~319--333.

\bibitem[Landis(2004)]{landis2004robots}
Landis, G.A.
\newblock Robots and humans: synergy in planetary exploration.
\newblock {\em Acta astronautica} {\bf 2004}, {\em 55},~985--990.

\bibitem[Pugh and Martinoli(2007)]{pugh2007multirobotpso}
Pugh, J.; Martinoli, A.
\newblock Inspiring and Modeling Multi-Robot Search with Particle Swarm
  Optimization.
\newblock  Swarm Intelligence Symposium, 2007. SIS 2007. IEEE,  2007, pp.
  332--339.

\bibitem[Zhu \em{et~al.}(2011)Zhu, Liang, and Guan]{zhu2011pso}
Zhu, Q.; Liang, A.; Guan, H.
\newblock A PSO-inspired multi-robot search algorithm independent of global
  information.
\newblock  Swarm Intelligence (SIS), 2011 IEEE Symposium on. IEEE,  2011, pp.
  1--7.

\bibitem[Sutantyo \em{et~al.}(2010)Sutantyo, Kernbach, Levi, Nepomnyashchikh,
  et~al.]{sutantyo2010multi}
Sutantyo, D.K.; Kernbach, S.; Levi, P.; Nepomnyashchikh, V.; others.
\newblock Multi-robot searching algorithm using l{\'e}vy flight and artificial
  potential field.
\newblock  Safety Security and Rescue Robotics (SSRR), 2010 IEEE International
  Workshop on. IEEE,  2010, pp. 1--6.

\bibitem[Baranzadeh and Savkin(2015)]{baranzadeh2015distributed}
Baranzadeh, A.; Savkin, A.V.
\newblock A distributed algorithm for grid-based search by a multi-robot
  system.
\newblock  Control Conference (ASCC), 2015 10th Asian. IEEE,  2015, pp. 1--6.

\bibitem[Cavalcante \em{et~al.}(2012)Cavalcante, Noronha, and
  Chaimowicz]{cavalcante2012local}
Cavalcante, R.C.; Noronha, T.F.; Chaimowicz, L.
\newblock A Local Search Approach for Improving Multi-Robot Routing in
  Exploration Missions.
\newblock  Robotics Symposium and Latin American Robotics Symposium (SBR-LARS),
  2012 Brazilian. IEEE,  2012, pp. 85--90.

\bibitem[Marjovi \em{et~al.}(2009)Marjovi, Nunes, Marques, and
  De~Almeida]{marjovi2009multi}
Marjovi, A.; Nunes, J.G.; Marques, L.; De~Almeida, A.
\newblock Multi-robot exploration and fire searching.
\newblock  Intelligent Robots and Systems, 2009. IROS 2009. IEEE/RSJ
  International Conference on. IEEE,  2009, pp. 1929--1934.

\bibitem[Burgard \em{et~al.}(2005)Burgard, Moors, Stachniss, and
  Schneider]{burgard2005coordinated}
Burgard, W.; Moors, M.; Stachniss, C.; Schneider, F.E.
\newblock Coordinated multi-robot exploration.
\newblock {\em IEEE Transactions on robotics} {\bf 2005}, {\em 21},~376--386.

\bibitem[Gelenbe(2010)]{gelenbe2010search}
Gelenbe, E.
\newblock Search in unknown random environments.
\newblock {\em Physical Review E} {\bf 2010}, {\em 82},~061112.

\bibitem[Lau and Ko(2007)]{lau2007immuno}
Lau, H.Y.; Ko, A.
\newblock An immuno robotic system for humanitarian search and rescue
  (application stream). In {\em Artificial Immune Systems}; Springer,  2007;
  pp. 191--203.

\bibitem[Su \em{et~al.}(2015)Su, Zhang, and Bai]{su2015dynamic}
Su, X.; Zhang, M.; Bai, Q.
\newblock Dynamic task allocation for heterogeneous agents in disaster
  environments under time, space and communication constraints.
\newblock {\em The Computer Journal} {\bf 2015}, {\em 58},~1776--1791.

\bibitem[Kang and M{\"u}ller(2012)]{kang2012sphere}
Kang, R.J.; M{\"u}ller, T.
\newblock Sphere and dot product representations of graphs.
\newblock {\em Discrete \& Computational Geometry} {\bf 2012}, {\em
  47},~548--568.

\bibitem[Gelenbe \em{et~al.}(2005)Gelenbe, Hussain, and
  Kaptan]{gelenbe2005simulating}
Gelenbe, E.; Hussain, K.; Kaptan, V.
\newblock Simulating autonomous agents in augmented reality.
\newblock {\em Journal of Systems and Software} {\bf 2005}, {\em 74},~255--268.

\bibitem[Nourbakhsh \em{et~al.}(2005)Nourbakhsh, Sycara, Koes, Yong, Lewis, and
  Burion]{nourbakhsh2005human}
Nourbakhsh, I.R.; Sycara, K.; Koes, M.; Yong, M.; Lewis, M.; Burion, S.
\newblock Human-robot teaming for search and rescue.
\newblock {\em Pervasive Computing, IEEE} {\bf 2005}, {\em 4},~72--79.

\bibitem[Gelenbe(2000)]{G-Nets}
Gelenbe, E.
\newblock The first decade of G-networks.
\newblock {\em European Journal of Operational Research} {\bf 2000}, {\em
  126},~231--232.

\bibitem[Aversa \em{et~al.}(2010)Aversa, Di~Martino, Rak, and
  Venticinque]{aversa2010cloud}
Aversa, R.; Di~Martino, B.; Rak, M.; Venticinque, S.
\newblock Cloud agency: A mobile agent based cloud system.
\newblock  Complex, Intelligent and Software Intensive Systems (CISIS), 2010
  International Conference on. IEEE,  2010, pp. 132--137.

\bibitem[Gelenbe and Loukas(2007)]{gelenbe2007self}
Gelenbe, E.; Loukas, G.
\newblock A self-aware approach to denial of service defence.
\newblock {\em Computer Networks} {\bf 2007}, {\em 51},~1299--1314.

\bibitem[Gelenbe(2012)]{gelenbe2012energy}
Gelenbe, E.
\newblock Energy packet networks: adaptive energy management for the cloud.
\newblock  Proceedings of the 2nd International Workshop on Cloud Computing
  Platforms. ACM,  2012, p.~1.

\bibitem[Gelenbe(2014{\natexlab{a}})]{gelenbe2014adaptive}
Gelenbe, E.
\newblock Adaptive management of energy packets.
\newblock  Computer Software and Applications Conference Workshops (COMPSACW),
  2014 IEEE 38th International. IEEE,  2014, pp. 1--6.

\bibitem[Gelenbe(2014{\natexlab{b}})]{gelenbe2014sensor}
Gelenbe, E.
\newblock A sensor node with energy harvesting.
\newblock {\em ACM SIGMETRICS Performance Evaluation Review} {\bf 2014}, {\em
  42},~37--39.

\end{thebibliography}




\end{document}